\documentclass[final,5p,times,twocolumn, authoryear]{elsarticle}

\usepackage{graphicx}          % Include this line if your 
\usepackage{natbib}
\usepackage[cmex10]{amsmath}
\usepackage{algorithmic}
\usepackage{array}
\usepackage{moreverb}
\usepackage{amsmath,amssymb}

\usepackage{subfigure}
\usepackage{bibentry, natbib}
\usepackage{soul, color}
\usepackage{indentfirst}
\usepackage{bm}
\usepackage{float}
\usepackage{stfloats}
\usepackage{bbm}
\usepackage{booktabs}

\usepackage[colorlinks,citecolor=darkblue,urlcolor=red,hyperfigures]{hyperref}

\newtheorem{theorem}{Theorem}
\newtheorem{lemma}{Lemma}

\newtheorem{remark}{Remark}
\newtheorem{assumption}{Assumption}
\newtheorem{problem}{Problem}
\newtheorem{proposition}{Proposition}

\renewcommand{\figurename}{Fig.}

\def \T{{\mbox{\tiny T}}}

\def\col{{\mbox{col}}}
\biboptions{sort&compress}
\journal{arXiv}
\begin{document}    

\begin{frontmatter}
%\runtitle{Insert a suggested running title}  % Running title for regular 
                                              % papers but only if the title  
                                              % is over 5 words. Running title 
                                              % is not shown in output.

\title{\LARGE 
% Generalized Discrete-Time Linear Quadratic Output Feedback Learning Control
% Generalized 
Optimal Output Feedback Learning Control for Discrete-Time Linear Quadratic Regulation
% Discrete-Time Linear Quadratic Output Feedback Learning Control
\tnoteref{footnoteinfo}} % Title, preferably not more 
                                                % than 10 words.

\tnotetext[footnoteinfo]{%Manuscript recieved \today. 
This work was supported
in part by the National Science and Technology Major Project under Grant 2021ZD0112600,
in part by the National Natural Science Foundation of China under Grant
61973035,
in part by Shanghai Municipal Science and Technology Major Project 2021SHZDZX0100,
in part by the Key Program of the National Natural Science Foundation of China under Grant 61933002,
in part by the National Science Fund for Distinguished Young Scholars of China under Grant 62025301,
in part by the Natural Science Foundation of Chongqing under Grant 2021ZX4100036,
in part by the Basic Science Center Programs of NSFC under Grant 62088101,
in part by the Postdoctoral Fellowship Program of CPSF Project GZC20233407. Corresponding author: Maobin Lu. }

% and in part by the Natural Science Foundation of Fujian Province of China under Grant 2021J01051

\author[bit,cq]{Kedi Xie}\ead{kedixie@bit.edu.cn}
\author[QU]{Martin Guay}\ead{martin.guay@queensu.ca}
\author[MIT]{Shimin Wang}%\ead{shimin.wang@queensu.ca}
\author[bit,cq]{Fang Deng}\ead{dengfang@bit.edu.cn}
\author[bit,cq]{Maobin Lu}\ead{lumaobin@bit.edu.cn}

\address[bit]{School of Automation, Beijing Institute of Technology, Beijing 100081, China.}

\address[QU]{Department of Chemical Engineering, Queen’s University, Kingston, ON K7L 3N6, Canada.}
\address[MIT]{Massachusetts Institute of Technology, Cambridge, MA 02139, USA.}

\address[cq]{ Beijing Institute of Technology Chongqing Innovation Center, Chongqing 401120, China.}

                             % keyword list or with the 
                                          % help of the Automatica 
                                          % keyword wizard

\begin{abstract}
This paper studies the linear quadratic regulation (LQR) problem of unknown discrete-time systems via dynamic output feedback learning control. 
In contrast to the state feedback, the optimality of the dynamic output feedback control for solving the LQR problem requires an implicit condition on the convergence of the state observer. 
Moreover, due to unknown system matrices and the existence of observer error, it is difficult to analyze the convergence and stability of most existing output feedback learning-based control methods. 
To tackle these issues, we propose a generalized dynamic output feedback learning control approach with guaranteed convergence, stability, and optimality performance for solving the LQR problem of unknown discrete-time linear systems. 
In particular, a dynamic output feedback controller is designed to be equivalent to a state feedback controller. This equivalence relationship is an inherent property without requiring convergence of the estimated state by the state observer, which plays a key role in establishing the off-policy learning control approaches.
By value iteration and policy iteration schemes, the adaptive dynamic programming based learning control approaches are developed to estimate the optimal feedback control gain.
%The equivalent relationship between the state feedback and dynamic output feedback plays a key role in establishing the off-policy learning control approaches.
%
In addition, a model-free stability criterion is provided by finding a nonsingular parameterization matrix, which contributes to establishing a switched iteration scheme.
Furthermore, the convergence, stability, and optimality analyses of the proposed output feedback learning control approaches are given. 
Finally, two examples are used to verify the effectiveness of the proposed dynamic output feedback learning control approach.
\end{abstract}

\end{frontmatter}

\section{Introduction}\label{sec-intro}

Reinforcement learning (RL) \citep{Sutton2018, Lewis2012}, also known as adaptive dynamic programming (ADP) \citep{Jiang2017_book, Powell2007, Liu2021, Lewis2012introduction} in the field of control, is one of the most important cutting-edge research fields in artificial intelligence for optimization \cite{wu2025eiqp} and decision-making in uncertain environments.
Especially, Reinforcement learning methods have shown promise in monitoring real-time green carbon dioxide tracking in refinery processes \cite{cao2025machine} and in active traffic management \cite{su2020neuro}.
%
%For example, parameter learning plays an important role in monitoring the health of Lithium-ion batteries as illustrated in \cite{qian2024fully,che2023opportunities}, with the estimation of temperature parameters significantly improving the accuracy of battery health monitoring, as demonstrated in \cite{che2023battery}.%
%
Moreover, the ADP algorithm offers iteration learning frameworks to estimate the optimal control policy for solving several optimal control problems by only accessing a series of historical data of input, output, and/or state instead of prior knowledge of system dynamics; see, \cite{vrabie2009adaptive, Jiang2012, Modares2016,   Chen2023, Modares2014, Gao2016_1tac, Yang2018_1, Rizvi2020_aut, Duan2023}, and references therein.
As an ongoing research topic, it is of great significance to advance the development of the ADP method with output feedback control, as it has the potential to effectively address the optimal control problem under the unmeasurable state in various practical settings.
However, it is shown in \cite{Duan2023, Rizvi2023TAC, Chen2023} that the theoretical guarantees of the ADP-based output feedback learning control methods, including convergence, stability, and optimality, remain substantial challenges.

As is widely known, the linear quadratic regulator is a full-state feedback controller, and its optimal control gain can be obtained by solving a corresponding algebraic Riccati equation (ARE) \citep{kalman1960contributions, Lancaster1995} which relies on accurate knowledge of the system matrices.
For the purpose of obtaining the optimal state feedback control policy without identifying system matrices, a novel ADP-based learning algorithm was first proposed in \cite{vrabie2009adaptive} to solve the LQR problem of partially unknown linear systems. 
Using a single state trajectory, a policy iteration (PI) scheme was employed in \cite{vrabie2009adaptive} to estimate the optimal control policy in an online fashion, and the convergence is guaranteed by requiring an initial stabilizing control policy and the persistence of excitation (PE) condition.
Later on, by using an integral RL technique with the available data information of state and input, an ADP-based off-policy model-free learning algorithm with the PI scheme was proposed in \cite{Jiang2012} to approximate the optimal state feedback controller for solving the LQR problem of completely unknown linear systems. 
To relax the requirement of initial admissible control policy, a value iteration (VI) based learning algorithm was proposed in \cite{Bian2016} at the cost of convergence rate. 
In addition, the stability analysis of the ADP-based state feedback learning control approach was provided in \cite{Bian2016}. It is worth noting that both the convergence and optimality of the proposed learning algorithms in \cite{Jiang2012} and \cite{Bian2016} can be guaranteed under a rank condition which is a milder condition than the PE condition. 
The ADP-based learning framework in \cite{vrabie2009adaptive, Jiang2012, Bian2016} for solving the LQR problem has been extended to solve several optimal control problems, for instance, the optimal tracking problem \citep{Modares2014}, the optimal robust control problem \citep{Bian-2022-robust}, and the optimal output regulation problem \citep{Gao2016_1tac,  Gao2020-NZS}.
These ADP-based state feedback learning control approaches provide the fundamental results for developing the output feedback results.

%To deal with the unmeasurable states, a stabilizing dynamic output feedback controller for linear systems can be constructed via a stable state observer and a stable state feedback controller, which is well-known as the separation principle. 
Based on the existing results on model-free state feedback \citep{vrabie2009adaptive, Jiang2012, Bian2016}, several output feedback learning control approaches were proposed by designing a stable state observer with parameterization methods, for instance, \cite{Lewis2011, Modares2016,   Chen2022, Chen2023, Rizvi2020_aut, Rizvi2023TAC, Gao2019, Gao2016-3aut}. 
In particular, a state parameterization method for linear discrete-time systems was first designed in \cite{Lewis2011}, which makes the output feedback controller with the constructed state equivalent to a state feedback controller. 
Similar to the state feedback learning control in \cite{vrabie2009adaptive}, by using the measurement of the output and input data with a probing noise to replace the system matrices in the corresponding ARE, a model-free learning control approach was proposed to solve the LQR problem in \cite{Lewis2011}. 
It should be pointed out that the probing noise would lead to an estimation error in the learning process.
The framework of output feedback control with the state parameterization method and the ADP algorithm was widely extended to solve several optimal control problems, for example, the optimal robust control problem \citep{Gao2016-3aut} and the optimal output regulation problem \citep{Gao2019}. 
%
%To overcome the issue of exploration noise, a modified cost function was designed in \cite{Lewis2011, Modares2016}, and an ADP-based learning algorithm was proposed to estimate the discounted controller. 
Nevertheless, as stated in \cite{postoyan2016stability}, the discounted controller in \cite{Lewis2011, Modares2016} is not an optimal solution and may not even be a stabilizing one if the discounting factor in the cost function is not carefully chosen.

Different from \cite{Lewis2011, Modares2016}, a novel state parameterization method was developed in \cite{rizvi2019DToutput} to solve the output feedback LQR problem of discrete-time linear systems. 
In particular, an internal system was embedded in the controller that can be seen as a stable state observer under the observability condition.
In addition, the problem of exploration noise bias was avoided in \cite{rizvi2019DToutput} without resorting to a discounting factor by proposing a model-free Q-learning output feedback control method with both PI and VI schemes. 
This output feedback learning control framework designed in \cite{rizvi2019DToutput} has been expanded to solve the tracking problems \citep{Chen2022, Chen2023} and the zero-sum games \citep{Rizvi2020_aut}. However, as stated in \cite{Rizvi2023TAC}, the convergence and optimality of the output feedback learning control algorithm in the aforementioned references \citep{rizvi2019DToutput, Rizvi2020_aut,   Chen2022, Chen2023} may not be guaranteed unless the convergence of the state observer is achieved. 
%The essential reason is that the output feedback learning control methods in these works are established based on the equivalence of the state feedback and output feedback learning equations.
Moreover, stability analysis has not been provided in these works, while it is a bare requirement in a closed-loop control design.

In this paper, we aim to give a generalized result of the ADP-based output feedback learning control for solving the discrete-time LQR problem.
The main significance of the result is that we provide an optimal solution to the output feedback LQR problem with convergence guaranteed, and the stability analysis is also given for the learning process.
%To relax the implicit condition in the convergence of the state observer, we present a generalized dynamic output feedback learning control method of which convergence and optimality can be guaranteed. Besides, we provide a model-free stability criterion to analyze the stability of the closed-loop systems under the estimated control policy at any iteration.
Compared to the recent works on ADP-based output feedback learning control \citep{rizvi2019DToutput, Rizvi2023TAC, Chen2022}, the contributions of this paper are threefold. 
\begin{itemize}
    \item[1)] A generalized dynamic output feedback learning controller is proposed, which has an inherent property of being equivalent to a state feedback controller.
    The equivalence relationship holds without the requirement of the convergence of state observer, which not only contributes to establishing the ADP-based learning algorithms but also ensures the optimality of the proposed dynamic output feedback learning controller.
    \item[2)] Two model-free ADP-based learning algorithms are proposed by employing the PI scheme and the VI scheme, respectively. 
    The convergence of the proposed learning algorithms with both PI and VI can be guaranteed under the rank conditions, and it is immune to the exploration noise and the state observer error. 
    In addition, the rank condition of state parameterization can be achieved with more flexible pole assignment configurations.
    \item[3)] A model-free stability criterion is provided, and the stability analysis of the learning algorithm is given. 
    The stability criterion paves the way for us to propose a switched iteration (SI) scheme integrated with PI and VI to solve the LQR problem. 
    The proposed SI scheme can relax the requirement of an initial admissible controller in the PI scheme and accelerate the convergence rate in the VI scheme.
\end{itemize}

The rest of this paper is organized as follows. Section \ref{sec-pre} presents the preliminaries and problem formulation. In Section \ref{sec-main}, a novel dynamic output feedback control design method is proposed, and the ADP-based learning control approach is developed. The stability, convergence, and optimality analysis are also presented in this section. Finally, the simulation results are given in Section \ref{sec-sim}, and Section \ref{sec-con} draws the conclusion.

\section{Preliminaries and Problem Formulation}\label{sec-pre}

%\emph{Notations:}
\subsection{Notation} \label{notations}
Throughout this paper, $\mathbb{R}$ and $\mathbb{Z}^+$ denote the sets of real numbers and non-negative integers, respectively. $\mathbb{U}^-$ denotes the set of complex numbers that are strictly inside the unit circle. For a square matrix $A$, $\sigma(A)$ is the complex spectrum of $A$. $A>0$, $A \geq 0$, $A < 0$, and $A \leq 0$ represent that $A$ is positive definite, positive semi-definite, negative definite, and negative semi-definite, respectively. For any given $n, m \in \mathbb{Z}^+$, $I_n \in \mathbb{R}^{n \times n} $ is a unit matrix, and $ \mathbf{0} $ is a zero vector or matrix with proper dimension. For column vectors $v_i \in \mathbb{R}^{n_i \times m}$, $i=1,2,\dots,N$, define $\mbox{col}(v_1,v_2,\dots,v_N)=[v_1^\T, v_2^\T, \dots, v_N^\T]^\T \in \mathbb{R}^{(n_1+n_2+\dots +n_N) \times m}$, and $\mbox{dia}(x)$ denotes a diagonal matrix $X=[X_{ij}] \in \mathbb{R}^{n \times n}$ with $X_{ij}=0,~i\neq j$ and $X_{ii}=x_i$. For any given matrix $B \in \mathbb{R}^{m\times n}$ and a symmetric matrix $C \in \mathbb{R}^{n\times n}$, define $\mbox{vec}(B)=[b_1^\T,b_2^\T,\dots,b_n^\T]^\T$ with $b_i \in \mathbb{R}^m$ being the columns of $B$, and $\mbox{vech}(C)=[c_{11},c_{12}, \dots, c_{1n},c_{22},c_{23}, \dots, c_{n-1,n},c_{nn}]^\T \in \mathbb{R}^{\frac{n(n+1)}{2}}$ with $c_{ij}$ being the element of matrix $C$. $B^{\dagger}$ denotes the pseudo-inverse of $B$. For any matrix $G_i$, $i=1,\dots,n$, $\mathcal{G}=\mbox{block diag} [G_1, \dots, G_n]$ is an augmented block diagonal matrix with block matrix $\mathcal{G}_{ii}={G_i}$. The symbol $\otimes$ denotes the Kronecker product. $\mathcal{Z}^{-1}(\cdot)$ represents the inverse $z$ transform operation. $\| \cdot \|$ is the induced norm for matrices. The functions $\mbox{det}(\cdot)$ and $\mbox{adj}(\cdot)$ denote the determinant and adjoint matrix.

\subsection{Plant and Preliminaries}

Consider the following linear discrete-time systems
\begin{subequations}\label{sys_origin}
\begin{align}
\label{sys_origin_x} x({k+1}) &= Ax(k) + B u (k)\\
\label{sys_origin_y} y(k)     &= Cx(k) %, ~~k=0, 1, 2,\dots   
\end{align}
\end{subequations}
where $x\in\mathbb{R}^{n}$, $u \in \mathbb{R}^{m}$, and $y \in\mathbb{R}^{p}$ are the \emph{unmeasurable state}, the control input, and the output, respectively. $A$, $B$, and $C$ are constant system matrices with appropriate dimensions and satisfy the following condition.
\begin{assumption} \label{ass_1}
The pair $(A, B)$ is stabilizable, and the pair $(A, C)$ is observable.  \hfill \rule{1.5mm}{1.5mm}
\end{assumption}

Define the utility function as
\begin{flalign} \label{cost_function}
\mathcal{C}_y(y(k),u(k))
=y^\T(k) {Q_y} y(k) + u^\T(k) {R} u (k)
\end{flalign}\noindent
where $Q_y = Q_y^\T \in \mathbb{R}^{p \times p} \geq \mathbf{0}$, $R = R^\T \in \mathbb{R}^{m \times m}> \mathbf{0}$ are the weight matrices, and the pair $(A, \sqrt{ C^\T Q_y C})$ is observable. The objective of the LQR problem is to find a feedback control sequence $u(k)$ such that the following infinite horizon quadratic value function
\begin{flalign} \label{value_function}
V(x(0))
=\sum_{k=0}^\infty  
y^\T(k) {Q_y} y(k) + u^\T(k) {R} u (k)
\end{flalign}
is minimized.

If the state $x(k)$ is measurable, the LQR problem can be solved by the well-known full-state feedback controller
\begin{flalign}\label{full-state-controller}
    u(k) = - {K^ * }{x(k)}
\end{flalign}
where the optimal control gain is
\begin{flalign}\label{ARE K}
    K^*=\left( R + B^\T  P^* B \right)^{-1} B^\T  P^* A
\end{flalign}
with the symmetric matrix $P^*$ as the solution to the discrete algebraic Riccati equation (ARE) \citep{Lancaster1995}
\begin{flalign}\label{ARE}
{A}^\T {P }A - P + C^\T Q_y C - A^\T PB \left( R + B^\T  P B \right)^{-1} B^\T  P A = \mathbf{0}.
\end{flalign}\noindent
Then, the value function \eqref{value_function} under the optimal controller \eqref{full-state-controller} is given as
\begin{flalign}
    V(x(0)) = x^\T(0) P^* x(0).
\end{flalign}

That is, the LQR problem can be solved by computing ARE \eqref{ARE} and \eqref{ARE K} when the state is available. Based on this result, it is not difficult to know that the LQR problem with unmeasurable state can be solved by reconstructing or estimating the real state according to the separation principle \citep{Lewis2012}. As given in \citet[Lemma 1]{Lewis2011}, under the observability condition, the state of system \eqref{sys_origin} can be reconstructed by
\begin{flalign}\label{state reconstruction}
    x(k) =& M_{\bar y} \bar y_{k-1, k-N} + M_{\bar u} \bar u_{k-1, k-N} 
         = \left[ M_{\bar y},~ M_{\bar u}  \right] \left[ 
        \begin{matrix}
            \bar y_{k-1, k-N} \\
        \bar u_{k-1, k-N}
        \end{matrix} \right] \nonumber \\
        := & \bar M \bar \upsilon_{k-1, k-N}
\end{flalign}
where $N$ is a positive integer which is greater than the observability index of system \eqref{sys_origin}, $M_{\bar y}= A^N \left( V_N^\T V_N \right)^{-1}  V_N^\T$, $M_{\bar u} = U_N - M_{\bar y} T_N$, and
\begin{flalign*}
     \bar{y}_{k-1, k-N} = &\col\left(
        y(k-1),
        \dots,
        y(k-N-1),
        y(k-N) \right) \\ \bar{u}_{k-1, k-N} =& \col\left(
        u(k-1),
        \dots,
        u(k-N-1),
        u(k-N) \right)
\end{flalign*}
with 
% $U_N = \left[ B,~AB,~\cdots,~A^{N-1}B \right]$ and
\begin{flalign*}
%\begin{align}
U_N =& \left[ B,~AB,~\cdots,~A^{N-1}B \right]\\
    V_N =& \left[ \begin{matrix}
        CA^{N-1} \\
        \vdots \\
        CA\\
        C
    \end{matrix} \right], ~~ T_N = \left[ \begin{matrix}
       \mathbf{0} & CB & CAB & \cdots   & CA^{N-2} B \\
       \mathbf{0} & \mathbf{0} & CB & \cdots  & CA^{N-3} B \\
       \vdots &  \vdots &  \vdots & \ddots  & \vdots \\
       \mathbf{0} & \mathbf{0} & \mathbf{0} & \cdots  & C B \\
       \mathbf{0} & \mathbf{0} & \mathbf{0} & \cdots  & \mathbf{0} \\
    \end{matrix} \right].
%    \end{align}
\end{flalign*}
Using the state reconstruction in \eqref{state reconstruction}, the LQR problem can be solved by the following optimal controller:
\begin{flalign}\label{u state reconstruction}
    u(k) = -K^* \bar M \bar \upsilon_{k-1, k-N} := -K_{\bar M} \bar \upsilon_{k-1, k-N}.
\end{flalign}

Different from the state reconstruction given in \eqref{state reconstruction} which is specific to the discrete-time systems, an alternative way to obtain the state information is to use a state observer to estimate the real state. Concretely, the state of system \eqref{sys_origin} can be estimated by using the following Luenberger observer:
\begin{flalign}\label{Luenberger observer}
    \hat x (k+1) = (A-LC) \hat x(k) + B u(k) + L y(k).
\end{flalign}
It is known that the Luenberger observer \eqref{Luenberger observer} has the property that $\lim_{k \rightarrow \infty} \hat x(k) = x(k)$ when all the eigenvalues of $(A-LC)$ are set within the unit circle under the observability condition.
As given in \cite{rizvi2019DToutput}, by setting $\hat x(k) = \bm{0}$, the estimated state $\hat x(k)$ can be parametrized by
\begin{flalign}\label{state parameterization}
   \hat x(k) = W_y \omega (k) + W_u \sigma (k) = \left[ W_{y},~ W_{u}  \right]  \left[ 
        \begin{matrix}
             \omega (k) \\
        \sigma (k)
        \end{matrix} \right]
        := \bar W \varpi (k)
\end{flalign}
where the parameterization matrices $W_y\in \mathbb{R}^{n \times np}$ and $W_u\in \mathbb{R}^{n \times nm}$ are determined by the coefficients of the numerators in the transfer function matrix of the Luenberger observer system \eqref{Luenberger observer} with $y(k)$ and $u(k)$ being the inputs to the observer system, and the dynamics of $\omega (k)$ and $\sigma (k)$ are given as
\begin{flalign*}
    \omega (k+1) =&~ {\bar {\mathcal{A}}}_y \omega (k) + {\bar {\mathcal{B}}}_y y(k),~ \omega(0) = \mathbf{0} \\
    \sigma (k+1) =&~ {\bar {\mathcal{A}}}_u \sigma (k) + {\bar {\mathcal{B}}}_u u(k),~ \sigma(0) = \mathbf{0}
\end{flalign*}
where ${\bar {\mathcal{A}}}_y \in \mathbb{R}^{np \times np}$ and ${\bar {\mathcal{A}}}_u \in \mathbb{R}^{nm \times nm}$ are user-defined matrices determined by the eigenvalues of $(A-LC)$, and ${\bar {\mathcal{B}}}_y \in \mathbb{R}^{np \times p}$ and ${\bar {\mathcal{B}}}_u \in \mathbb{R}^{nm \times m}$ are also user-defined matrices which can make $({\bar {\mathcal{A}}}_y , ~{\bar {\mathcal{B}}}_y)$ and $({\bar {\mathcal{A}}}_u , ~{\bar {\mathcal{B}}}_u)$ controllable. It is easy to see that the data of $\omega(k)$ and $\sigma(k)$ are available since their system matrices $({\bar {\mathcal{A}}}_y , ~{\bar {\mathcal{B}}}_y)$ and $({\bar {\mathcal{A}}}_u , ~{\bar {\mathcal{B}}}_u)$ are user-defined.

Using the state parameterization in \eqref{state parameterization}, the LQR problem can be solved by
\begin{flalign}\label{u state parameterization}
    u(k) = -K^* \bar W \varpi (k)  := -K_{\bar W} \varpi (k)
\end{flalign}
when the state of the observer system converges to the real state. It is worth mentioning that, if all the eigenvalues of $(A-LC)$ are chosen as $0$, the observer state $\hat x(k)$ converges to the real state $x(k)$ when $k \geq n$.

Thus, by using the state reconstruction in \eqref{state reconstruction} or the state parameterization in \eqref{state parameterization}, the LQR problem can be solved by the output feedback controller \eqref{u state reconstruction} or \eqref{u state parameterization} when the accurate knowledge of system matrices are known.

\subsection{Problem Formulation}

In this paper, we aim to solve the LQR problem of linear discrete-time systems with completely unknown system matrices and unmeasurable state, which can be formulated as follows:
\begin{problem} \label{problem 1}
    For system \eqref{sys_origin} where the system matrices $A$, $B$, and $C$ are unknown, and the state $x$ is unmeasurable, find an optimal control policy sequence $u(k)$ to satisfy
    \begin{flalign*}
        V(x(0)) = \sum_{k=0}^\infty  
y^\T(k) {Q_y} y(k) + u^\T(k) {R} u (k) = x^\T (0) P^* x(0)
    \end{flalign*}
    where $P^*$ is the solution to ARE \eqref{ARE}. \hfill \rule{1.5mm}{1.5mm}
\end{problem}

The challenges of Problem \ref{problem 1} arise from the solution of ARE \eqref{ARE} and the design of the parameterization matrix $\bar M$ or $\bar W$, which becomes difficult in the absence of knowledge of the system matrices $A$, $B$, and $C$.

By combining the ADP method with a series of historical data of input and output, some ADP-based output feedback learning approaches focused on estimating the optimal control gain $K_{\bar M}$ and $K_{\bar W}$ directly, for instance, \cite{Lewis2011, rizvi2019DToutput, Rizvi2020_aut, Rizvi2023TAC, Gao2016-3aut, Chen2023}. Particularly, it is proven in \cite{Rizvi2023TAC} that the requirement on the full row rank of $\bar W$ is necessary to guarantee the convergence performance of the aforementioned ADP-based output feedback learning approaches, so is the same requirement on $\bar M$ along the similar proof.
However, as we can see from \eqref{state reconstruction}, under Assumption 1, the parameterization matrix $\bar M$ may not be of full row rank when the system matrix $A$ has the eigenvalue $0$. It follows from \citet[Theorem 4]{Rizvi2023TAC} that the parameterization matrix $\bar W$ may not be of full row rank when the matrices $(A-LC)$ and $A$ have common eigenvalues.
%the observer matrix $(A-LC)$ has the same eigenvalues as the system matrix $A$. 
On the other hand, as stated in \cite{postoyan2016stability}, the stability of the closed-loop system under the learning control approach proposed in \cite{Lewis2011} cannot be guaranteed. Although this issue was eliminated in \cite{rizvi2019DToutput, Rizvi2023TAC, Chen2023}, the requirement of the convergence of state observer is needed to ensure the equivalence of the dynamic output feedback controller with the state feedback controller. This implies that the observer error will influence the convergence and optimality performance of the ADP-based output feedback learning control approaches in \cite{rizvi2019DToutput, Rizvi2023TAC, Chen2023}.

To deal with the above issues, the objective of this paper is to provide a generalized learning based output feedback solution to Problem \ref{problem 1} and a detailed analysis of convergence, stability, and optimality for the proposed output feedback learning control approach.

\section{Main Results}\label{sec-main}

In this section, we propose a novel output feedback learning control approach to solve the LQR problem. In particular, a new dynamic output feedback controller is designed that is equal to a static state feedback controller. Then, a data-driven learning algorithm is established to estimate the optimal control gain without prior knowledge of system matrices. Finally, the convergence, stability, and optimality analyses of the proposed learning control approach are given.

\subsection{Dynamic Output Feedback Controller Design}\label{sec-control design}

Now we present a generalized dynamic output feedback controller design method, where the internal model is established without using any prior knowledge of the system dynamics. Moreover, the equivalence relationship between the proposed output feedback controller and the state feedback controller always holds even in the presence of observer error.

\begin{theorem} \label{theorem control design}
Under Assumption \ref{ass_1}, there exists a dynamic output feedback controller
\begin{subequations}\label{controller MF}
\begin{flalign} 
\label{controller MF u}  u(k)   &= -\mathcal{K} \eta (k)  \\
\label{controller MF z}  \eta (k+1) &=  \mathcal{G}_1 \eta (k) + \mathcal{G}_2 \zeta(k)
\end{flalign}
\end{subequations}
that is equal to a static state feedback controller $u(k)=-Kx(k)$, $\forall k = 0, 1, 2, \dots$, 
where $\eta \in \mathbb{R}^{n(m+p+1)}$ is the state of the internal model \eqref{controller MF z} with a user-defined matrix pair $(\mathcal{G}_1, \mathcal{G}_2)$, $\mathcal{K}=KM$ is the control gain with a parameterization constant matrix $M$, and $\zeta =\textnormal{\col}(u,y,0)$.
\end{theorem}

\emph{Proof.} Consider a Luenberger observer system described as in \eqref{Luenberger observer} and rewiritten as
\begin{flalign}\label{sys_observer}
    \hat x(k+1) = A_L \hat x(k) +B u(k) + Ly(k), ~~\hat x(0) = 0
\end{flalign}
where $A_L=A-LC$ is the observer system matrix with observer gain $L$. By Assumption \ref{ass_1}, the characteristic polynomial of $A_L$ can be designed arbitrarily, i.e.,
\begin{flalign}\label{characteristic polynomial}
    \Lambda(z) \!=\! \mbox{det}(z I_n \!-\!A_L)  \!=\! z^{n}\!+\!\alpha_{n\!-\!1} z^{n\!-\!1} \!+\!\cdots \!+\! \alpha_1 z \!+\! \alpha_0
\end{flalign}
is user-defined with known coefficients $\alpha_0, \alpha_1, \dots, \alpha_{n-1}$. 

Define the error as 
\begin{flalign}\label{error_define}
    \varepsilon_x(k) = x(k) - \hat x (k).
\end{flalign}
If follows from \eqref{sys_origin} and \eqref{sys_observer} that the dynamics of error is
\begin{flalign}\label{sys_error}
    \varepsilon_ x(k+1) = A_L \varepsilon_x(k).
\end{flalign}
Using the $z$-transformation, the explicit solution to $\varepsilon_x(k)$ in \eqref{sys_error} is given as
\begin{align}\label{derivation z transformation}
   \varepsilon_x(k) &= \mathcal{Z}^{-1}\left( (z I_n-A_L)^{-1}{\varepsilon_x (0)} \right) \nonumber \\
    &= \mathcal{Z}^{-1}\left( \frac{\mbox{adj}(z I_n-A_L)\varepsilon_x (0)}{\mbox{det}(z I_n-A_L)} \right)  \nonumber \\  
    =& \mathcal{Z}^{-1}\left( \left[ \begin{matrix}
        \frac{a^{1}_{n-1} z^{n-1} + a^{1}_{n-2} z^{n-2} + \cdots+a^{1}_1 z +a^{1}_0}{z^{n}+\alpha_{n-1} z^{n-1} +\cdots+\alpha_1 z +\alpha_0} \\
        \frac{a^{2}_{n-1} z^{n-1} + a^{2}_{n-2} z^{n-2} + \cdots+a^{2}_1 z +a^{2}_0}{z^{n}+\alpha_{n-1} z^{n-1} +\cdots+\alpha_1 z +\alpha_0}\\
        \vdots \\
        \frac{a^{n}_{n-1} z^{n-1} + a^{n}_{n-2} z^{n-2} + \cdots+a^{n}_1 z +a^{n}_0}{z^{n}+\alpha_{n-1} z^{n-1} +\cdots+\alpha_1 z +\alpha_0} \\
    \end{matrix} \right]   \right)  \nonumber \\
        =& \mathcal{Z}^{-1}\left( \left[ \begin{matrix}
        a^{1}_{0} & a^{1}_{1} & \cdots  & a^{1}_{n-1} \\
        a^{2}_{0} & a^{2}_{1} & \cdots & a^{2}_{n-1} \\
        \vdots    &  \vdots   & \cdots & \vdots\\
        a^{n}_{0} & a^{n}_{1} & \cdots & a^{n}_{n-1} \\
    \end{matrix} \right] \left[\begin{matrix}
        \frac{1}{\Lambda(z)}\\
        \frac{z}{\Lambda(z)}\\
        \vdots \\
        \frac{z^{n-1}}{\Lambda(z)}\\
    \end{matrix}  \right] \right)  \nonumber \\
    :=& M_{\varepsilon_x}(A,L,C,\varepsilon_{0}) \mathcal{Z}^{-1}\left( \left[\begin{matrix}
        \frac{1}{\Lambda(z)}\\
        \frac{z}{\Lambda(z)}\\
        \vdots \\
        \frac{z^{n-1}}{\Lambda(z)}\\
    \end{matrix}  \right] \right) 
\end{align}
where $M_{\varepsilon_x}(A,L,C,\varepsilon_x (0))$ is an unknown constant matrix determined by $(A,L,C,\varepsilon_x (0))$. 

Define a new system described by
\begin{flalign*}
    \eta_\varepsilon (k+1) = \mathcal{A}_\varepsilon \eta_\varepsilon (k)
\end{flalign*}
where the system matrix $\mathcal{A}_\varepsilon$ is known %which 
and has the same polynomial as $A_L$. Along the same derivation as in \eqref{derivation z transformation}, the explicit solution to $\eta_{\varepsilon} (k)$ is
\begin{flalign*}
    \eta_{\varepsilon}(k)  = M_{\eta_{\varepsilon}}(\mathcal{A}_\varepsilon,\eta_{\varepsilon} (0) ) \mathcal{Z}^{-1}\left( \left[\begin{matrix}
        \frac{1}{\Lambda(z)}\\
        \frac{z}{\Lambda(z)}\\
        \vdots \\
        \frac{z^{n-1}}{\Lambda(z)}
    \end{matrix}  \right] \right)
\end{flalign*}
where $M_{\eta_{\varepsilon}}(\mathcal{A}_\varepsilon,\eta_{\varepsilon} (0))$ is a computable constant matrix determined by the user-defined $\mathcal{A}_\varepsilon$ and $\eta_{\varepsilon} (0)$. 
It is easy to choose $\eta_{\varepsilon} (0)$ satisfying $M_{z_{\varepsilon}}(\mathcal{A}_\varepsilon, \eta_{\varepsilon} (0))$ of full rank. Thus, we have
\begin{flalign} \label{virtual observer error transformation}
	\varepsilon_x (k) = M_{\varepsilon_x} M_{\eta_{\varepsilon}}^{-1} \eta_{\varepsilon} (k) := M_{\varepsilon_x \leftarrow \eta_{\varepsilon}} \eta_{\varepsilon}(k), ~~~\forall k = 0, 1, 2, \dots
\end{flalign}\noindent
where $M_{\varepsilon_x \leftarrow \eta_\varepsilon}$ is an unknown parameterization transformation matrix.

Along the similar derivation in \cite{Rizvi2023TAC}, using the $z$-transformation, the explicit solution to $\hat x(k)$ in \eqref{sys_observer} is given as
\begin{align} \label{transformation x hat x}
\hat {x} (k) & = \mathcal{Z}^{-1}\left( \sum_{i=1}^{m} {(zI_n-A_L)}^{-1} {B_i U_i(z)} + \sum_{j=1}^{p} {(zI_n-A_L)}^{-1} {L_j Y_j(z)}\right) \nonumber \\
            &=\sum_{i=1}^{m} M_u^i \eta _u^i(k) + \sum_{j=1}^{p} M_y^j \eta _y^j(k)
\end{align}\noindent
where $M_u^i \in \mathbb{R} ^ {n \times {n}}$ and $M_y^j \in \mathbb{R} ^ {n \times {n}}$ are unknown constant matrices determined by the matrices $A$, $L$, $C$, and $B$, and $\eta _u^i(k)$ and $\eta _y^j(k)$ are the internal states with the dynamics as follows:
\begin{flalign*}
\eta _u^i (k+1) &= \mathcal{A} z_u^i (k) + b u_i (k) \\
 \eta _y^j (k+1) &= \mathcal{A} z_y^j (k) + b y_j (k)
\end{flalign*}
with
\begin{flalign}\label{A_and_b}
\mathcal{A}=  \left[\!\!
                 \begin{array}{ccccc}
                   0 & 1 & 0 & \cdots & 0 \\
                   0 & 0 & 1 & \cdots & 0 \\
                  \vdots & \vdots & \vdots & \ddots & \vdots \\
                   0 & 0 & 0 & \cdots &1 \\
                   -\alpha_0 & -\alpha_1 & -\alpha_2 & \cdots & -\alpha_{n-1}
                 \end{array}
               \!\!\right],~
b= \left[\!\! \begin{array}{c}
      0 \\
      0 \\
      \vdots \\
      0 \\
      1
    \end{array} \!\!\right]
\end{flalign}\noindent
where $\mathcal{A}$ is determined by the coefficients of $\Lambda(s)$ as in \eqref{characteristic polynomial}. 

Combining \eqref{error_define}, \eqref{virtual observer error transformation}, and \eqref{transformation x hat x}, we have
\begin{flalign}
    x(k) &= \hat x(k) + \varepsilon_ x(k) \nonumber \\
         &=  \sum_{i=1}^{m} M_u^i \eta_u^i(k) + \sum_{j=1}^{p} M_y^j \eta_y^j(k) + M_{\varepsilon_x \leftarrow \eta_\varepsilon} \eta_{\varepsilon}(k) \nonumber \\
\label{transformation x eta}	     &:= M \eta(k) , ~~~~~~~~~~~k=0,1,\dots 
\end{flalign}
where 
\begin{flalign}
\label{M define}    M&\!=\![M_u^1, \dots, M_u^m, M_y^1, \dots, M_y^p, M_{\varepsilon_x \leftarrow \eta_\varepsilon}] \! \in \! \mathbb{R}^{n \times n_z} \\
\label{eta degine}    \eta &=\col \left(\eta_u^1, \dots, z_u^m, z_y^1, \dots, z_y^p, \eta_\varepsilon \right) \in \mathbb{R}^{n_z} 
\end{flalign}
with $n_z=n(m+p+1)$. 

Finally, the dynamic output feedback controller is designed as
\begin{subequations} \label{u_OPFB_z}
\begin{flalign} 
  u(k)   &= - \mathcal{K} \eta (k)  \\
\label{z_all_defined 2}  \eta (k+1) &=  \mathcal{G}_1 \eta (k) + \mathcal{G}_2 \zeta(k)
\end{flalign}
\end{subequations}
where $\mathcal{K} = KM$ is the control gain with $M$ defined in \eqref{M define}, $\zeta=\col(u,y,0)$, and
%\begin{subequations}\label{G define}
\begin{flalign}
    \label{G define} \mathcal{G}_1 = \mbox{block diag} [ \underbrace{\mathcal{A}, \cdots, \mathcal{A}}_{m+p}, \mathcal{A}_\varepsilon],  ~\mathcal{G}_2 = \mbox{block diag} [ \underbrace{b, \cdots, b}_{m+p}, \mathbf{0}].
\end{flalign} 
%\end{subequations}

It follows from \eqref{transformation x eta} that the dynamic output feedback controller \eqref{u_OPFB_z} is equal to the static state feedback controller $u(k)=-Kx(k)$, for all $k=0, 1, 2, \dots$. Thus, the proof is completed. \hfill \rule{1.5mm}{1.5mm}

Combining Theorem \ref{theorem control design} with the well-known linear quadratic regulator \eqref{full-state-controller}, the optimal dynamic output feedback controller can be given as
\begin{subequations} \label{u optimal MF}
\begin{flalign} 
  u(k)   &= - \mathcal{K}^* \eta (k)  \\
  \eta (k+1) &=  \mathcal{G}_1 \eta (k) + \mathcal{G}_2 \zeta(k)
\end{flalign}
\end{subequations}
with $\mathcal{K}^* = K^* M$.

\subsection{Output Feedback Learning Control Approach}
\label{sec-data-driven}

In this subsection, by only accessing the data information of input and output, we propose ADP-based learning algorithms to approximate the optimal feedback control gain $\mathcal{K}^*$, such that the optimality solution to the output feedback LQR problem, i.e., Problem \ref{problem 1}, can be obtained. 
%In particular, we establish a data-based learning equation to estimate the unknown parameters related to the value function. 
By combining the learning equation with the value iteration (VI) scheme and the policy iteration (PI) scheme, two model-free output feedback learning control approaches are proposed without any prior knowledge of system matrices.

To begin with, the well-known Bellman equation of the linear discrete-time systems can be written as
\begin{flalign}\label{bellman equation}
    V_j(x(k))
= V_j(x(k+1))  + \mathcal{C}_{x,u}(x(k), u(k)) 
\end{flalign}
where $V_j(x(k)) = x^\T (k)  P_j x (k)$ is the value function under the control policy $u(k)=-K_j x(k)$ and the utility function $\mathcal{C}_{x,u}(x(k),u(k))= x^\T (k) Q_x x(k) + u^\T(k) R u(k)$ with the weight matrix $Q_x=Q_x^\T \geq 0 \in \mathbb{R}^{n \times n}$ and $R=R^\T > 0 \in \mathbb{R}^{m \times m}$. 

The model-based VI scheme and model-based PI scheme for estimating the optimal feedback control gain $K^*$ are recalled in Algorithms 1 and 2, respectively. Note that the utility function $\mathcal{C}_{y}(y(k),u(k))$ in \eqref{cost_function} is a special case of $\mathcal{C}_{x,u}(x(k),u(k))$ by setting $Q_x = C^\T Q_y C$.

Based on the Bellman equation \eqref{bellman equation}, we next give an equivalent Bellman equation with the available state $\eta(k)$ of the internal model.

Along the system trajectories evaluated by \eqref{sys_origin}, one can obtain \begin{subequations}\label{VI_model_free x} \begin{flalign}
&x^\T(k)  P_j x (k)  \label{VI_model_free x eq1} \\
=& x ^\T(k\!+\!1)  P_j x (k\!+\!1)  + \mathcal{C}_{x,u}(x(k), u(k))  \label{VI_model_free x eq2} \\
= &\! \left[\!\! \begin{array}{cc}
		x (k)\\
		u (k) 
	\end{array} \!\! \right]^\T \! \left[\!\! \begin{array}{cc}
A^\T  P_j A \!+\!  Q_x  &   A^\T  P_j B  \\
B^\T  P_j A  &B^\T  P_j B \!+\! R
	\end{array} \!\! \right] \! \left[ \!\! \begin{array}{cc}
		x (k)\\
		u (k)
	\end{array} \!\! \right]. \label{VI_model_free x eq3}
\end{flalign}
\end{subequations}

For convenience of description, with the operators  in \emph{Notation}, define the functions $\delta_{v,w}(k)$ and $\delta_{v}(k)$ associated with the column vectors $v(k)$ and $w(k)$ as  
\begin{subequations}\label{operations}
    \begin{align}
         \delta_{v,w}(k) =& \left( w(k) \otimes v(k) \right)^\T \\
    \delta_{v}(k) =& \mbox{vech}(2{v}(k){v}{^\T }(k) -\mbox{dia}({v}(k))^2)^\T.
    \end{align}
\end{subequations}
Define the data regressor functions $\Delta_v(k_N)$, $\bar \Delta_v(k_N)$ and $\Gamma_{v,w} (k_N)$ associated with the column vectors sequences $v(k)$ and $w(k)$, $k=0, 1, \dots, k_N$, as 
\begin{subequations}\label{operations regressor}
    \begin{align}
    {\Delta _{v}(k_N)} =& [\delta_v(0)^\T, \delta_v(1)^\T, \dots, \delta_v(k_N)^\T] \\
    \bar \Delta_v(k_N) =& [\bar \delta_v(0)^\T, \bar \delta_v(1)^\T, \dots, \bar \delta_v(k_N)^\T] \\
    \Gamma_{v,w} (k_N) =& [\delta_{v,w}(0)^\T, \delta_{v,w}(1)^\T, \dots, \delta_{v,w}(k_N)^\T]
    \end{align}
\end{subequations}
with $\bar \delta_v (k) = \delta_v(k+1) - \delta_v(k)$.

\begin{table} \small
\begin{tabular} {p{0.9\columnwidth}}
\toprule
\emph{Algorithm 1.} Model-Based VI Scheme\\
\midrule
\textbf{Initialization.} %Give the weight matrices $Q_x=Q_x^T \geq 0$ and $R=R^\T >0$ satisfying $(A,\sqrt{Q_x})$ is observable. 
Choose an initial positive semi-definite matrix $P_0$ and a small threshold $\epsilon>0$. Set $j=0$.\\
%\textbf{Iteration Scheme.} Repeat the two-step iteration until convergence is met. \\
\textbf{loop:} \\
\qquad Update $P_{j+1}$ from 
\begin{flalign}
  \label{VI_model_based}   P_{j+1} =  A^\T  P_{j}  A + Q_x - A^\T  P_j  B \left( R + B^\T  P_j B \right)^{-1} B^\T  P_j A 
%\label{VI_model_based_K}    K_{j+1}=\left( R + B^\T  P_{j+1} B \right)^{-1} B^\T  P_{j+1} A ~~~~~~~~~~~~~~~~~~~~~~~~~~~
\end{flalign}
\qquad \textbf{if } $ \| {P}_{j+1} - P_{j}\| < \epsilon$ \\
\qquad \qquad \textbf{return} $\hat K=\left( R + B^\T  P_{j} B \right)^{-1} B^\T  P_{j} A$ \\
\qquad \textbf{end if} \\
\qquad $j = j+1 $ \\
\textbf{end loop}\\
\bottomrule
\end{tabular}
\end{table}

\begin{table} \small
\begin{tabular} {p{0.9\columnwidth}}
\toprule
\emph{Algorithm 2.} Model-Based PI Scheme\\
\midrule
\textbf{Initialization.} %Give the weight matrices $Q_x=Q_x^T \geq 0$ and $R=R^\T >0$ satisfying $(A,\sqrt{Q_x})$ is observable. 
Choose an initial admissible control gain $K_0$ and a small threshold $\epsilon>0$. Set $j=0$.\\
%\textbf{Iteration Scheme.} Repeat the two-step iteration until convergence is met. \\
\textbf{loop:} \\
\qquad Solve $P_j$ from 
\begin{flalign}
 \label{PI_model_based}    \mathbf{0} =  (A-BK_j)^\T  P_{j}  (A-BK_j) - P_j + Q_x + K_j^\T R K_j 
 \end{flalign}
\qquad \textbf{if } $j \geq 1$ \textbf{and} $ \| {P}_{j} - P_{j-1}\| < \epsilon$ \\
\qquad \qquad \textbf{return} $\hat K=K_j$ \\
\qquad \textbf{else } \\
 \qquad  \qquad Update $K_{j+1}$ from
 \begin{flalign}
\label{PI_model_based_K}    K_{j+1}=\left( R + B^\T  P_j B \right)^{-1} B^\T  P_j A 
\end{flalign}
\qquad \textbf{end if} \\
\qquad    $j = j+1 $ \\
\textbf{end loop}\\
\bottomrule
\end{tabular}
\end{table}

By the equivalence relationship given in \eqref{transformation x eta}, substituting the collected data vectors $\eta(k)$ and $u(k)$ into the functions $\delta_{v, w}$ and $\delta_{v}$, \eqref{VI_model_free x} can be rewritten as 
\begin{subequations} \label{VI_model_free z}
\begin{flalign}
& \eta ^\T(k)  \mathcal{P}_j \eta (k) \label{VI_model_free z eq0} \\
=& \eta ^\T(k+1)  \mathcal{P}_j \eta (k+1) + \mathcal{C}_{\eta,u}(\eta(k),u(k)) \label{VI_model_free z eq1} \\
= & \varphi (k) \left[\begin{matrix}
									\mbox{vech}(\mathcal{P}_{\eta,\eta}^{(j)})\\
									\mbox{vec}({\mathcal{P}}_{\eta,u}^{(j)}) \\
									\mbox{vech}({\mathcal{P}}_{u,u}^{(j)})
									\end{matrix} \right] \label{VI_model_free z eq2}
\end{flalign}
\end{subequations}
where
\begin{flalign*}
    {\mathcal{P}}_j =& M^\T P_j M \\
    \mathcal{P}_{\eta,\eta}^{(j)} =& M^\T (A^\T  P_{j} A+ Q_x) M\\
    \mathcal{P}_{\eta,u}^{(j)} =& M^\T A^\T  P_{j} B \\
    \mathcal{P}_{u,u}^{(j)} =& (B^\T  P_{j} B+R) 
\end{flalign*} 
and the data regressors are 
\begin{flalign*}
    \varphi (k) =& \left[ {\delta _{\eta}}(k),~ {2{\delta _{{\eta},{Ru}}}(k)},~ {\delta _{u}}(k)\right] \\
    \mathcal{C}_{\eta,u}(\eta(k),u(k)) =& \eta^\T(k) M^\T Q_x M \eta(k) + u^\T(k) R u(k).
\end{flalign*}
Under the utility function defined in \eqref{cost_function} which is the case of $Q_x = C^\T Q_y C$, the data regressor $\mathcal{C}_{\eta,u}(\eta(k),u(k))$ is calculable and given by
\begin{flalign}
    \mathcal{C}_{\eta,u}(\eta(k),u(k)) \triangleq & \mathcal{C}_y(y(k),u(k)) \nonumber \\
    = & y^\T(k) Q_y y(k) + u^\T(k) R u(k).
\end{flalign}

Based on the data-based Bellman equation \eqref{VI_model_free z}, the model-based VI scheme (Algorithm 1), and the model-based PI scheme (Algorithm 2), we next establish the model-free VI scheme and model-free PI scheme for solving Problem \ref{problem 1}.

\subsubsection{Model-Free VI Scheme}
\ 
\newline
\indent
Given an initial symmetric positive definite matrix $\mathcal{P}_0 \in \mathbb{R}^{n_z \times n_z}$, for each iteration $j = 0, 1, 2,  \dots$, the equality between \eqref{VI_model_free z eq1} and \eqref{VI_model_free z eq2} always holds for any utility function $\mathcal{C}_{\eta,u}(\eta(k),u(k))$. Thus, the vector-formed data-based learning equation is given as
\begin{equation} \label{VI ILE vector}
	\varphi (k) \left[\begin{matrix}
									\mbox{vech}(\mathcal{P}_{\eta,\eta}^{(j)})\\
									\mbox{vec}({\mathcal{P}}_{\eta,u}^{(j)}) \\
									\mbox{vech}({\mathcal{P}}_{u,u}^{(j)})
									\end{matrix} \right] = \psi (k) \mbox{vech} \left( {\mathcal{P}}_j \right)+ \mathcal{C}(k)
\end{equation}
where $\psi(k) \!=\! {\delta _{\eta}}(k+1)$ and $\mathcal{C}(k)= \mathcal{C}_{\eta,u}(\eta(k),u(k))$ is the value of utility function at instant $k$.

By the least squares implementation, the matrix-formed data-based learning equation used to estimate the unknown parameters $\left( \mathcal{P}_{\eta,\eta}^{(j)}, ~\mathcal{P}_{\eta,u}^{(j)}, ~\mathcal{P}_{u,u}^{(j)} \right)$ is given as
\begin{flalign}\label{VI ILE matrix}
\Pi (k_N)   \left[\begin{matrix}
									\mbox{vech}(\mathcal{P}_{\eta,\eta}^{(j)})\\
									\mbox{vec}({\mathcal{P}}_{\eta,u}^{(j)}) \\
									\mbox{vech}({\mathcal{P}}_{u,u}^{(j)})
									\end{matrix} \right]  = \Psi (k_N)   \mbox{vech} \left( {\mathcal{P}}_j \right)+\mathcal{L}(k_N)
\end{flalign}\noindent
where $\Pi (k_N) = \col \left( \varphi(0), \varphi(1), \dots, \varphi(k_N) \right) \in \mathbb{R}^{N \times n _{\varphi}}$ and $\Psi (k_N)  = \col \left( \psi(0), \psi(1),\dots, \psi (k_N) \right) \in \mathbb{R}^{N \times n _{\psi}}$ with dimensions $n _{\varphi} = \frac{(n_z+m)(n_z+m+1)}{2}$ and $n _{\psi}=\frac{n_z(n_z+1)}{2}$, and $\mathcal{L}(k_N)=\col \left( \mathcal{C}\left( 0 \right), \mathcal{C}\left( 1 \right), \dots, \mathcal{C}\left( k_N \right) \right)$. The matrices $\Pi (k_N)$, $\Psi (k_N)$, and $\mathcal{L}(k_N)$ are the data regressors during the data collection process $k \in [k_0, k_N]$. 

Combining the model-based VI scheme with the solution to \eqref{VI ILE matrix}, 
the value evaluation equation is given by
\begin{flalign}\label{P_VI_model_free}
\mathcal{P}_{j+1} \!=\! \mathcal{P}_{\eta,\eta}^{(j)} \!-\! \mathcal{P}_{\eta,u}^{(j)} \left(\mathcal{P}_{u,u}^{(j)} \right)^{-1} \left(\mathcal{P}_{\eta,u}^{(j)}\right)^\T.
\end{flalign}
The control policy can be estimated by
\begin{flalign}\label{K_VI_model_free}
	\mathcal{K}_{j} = \left( \mathcal{P}_{u,u}^{(j)} \right)^{-1} \left(\mathcal{P}_{\eta,u}^{(j)}\right)^\T.
\end{flalign}
Before we present the model-free learning algorithms with the VI scheme, we give a rank condition that is used to determine whether the collected data is enough for implementing the learning process.

\begin{theorem} \label{theorem_rank condition VI}
If there exists an instant $k_{N}^*$ such that the rank condition 
\begin{flalign}\label{rank condition}
\textnormal{rank}\left( \Pi (k_N^*) \right) = \frac{(n_z + m)(n_z + m+1)}{2}
\end{flalign}\noindent
is satisfied, then, for every iteration $j = 0, 1, 2, \dots$, the VI-based learning equation \eqref{VI ILE matrix} has a unique solution.
\end{theorem}

\emph{Proof.} If the rank condition \eqref{rank condition} is satisfied at $k_{N^*}$, then $\Pi (k_N^*)$ in \eqref{VI ILE matrix} is of full column rank at each iteration $j$, since $\Pi (k_N^*)$ only relies on the collected data and is independent on $\mathcal{P}_{j}$ and $\mathcal{K}_{j}$. Thus, there always exists a unique solution to the VI-based learning equation \eqref{VI ILE matrix} for every iteration $j$. \hfill \rule{1.5mm}{1.5mm}

Finally, the proposed output feedback learning control approach with the VI scheme can be summarized in Algorithm 3.

\begin{table}[bpt] \small
\begin{tabular} {p{0.95\columnwidth}}
\toprule
\emph{Algorithm 3.} Model-Free VI-Based Learning Control\\
\midrule
\textbf{Initialization.}  %Give the weight matrices $Q_y=Q_y^T \geq 0$ and $R=R^\T >0$ satisfying $(A,\sqrt{C^\T Q_y C})$ is observable. 
Choose a symmetric positive semi-definite matrix $\mathcal{P}_0$. Select a pair $(\mathcal{G}_1, \mathcal{G}_2)$, an initial internal state $\eta(0)$, an initial control gain ${\mathcal{K}_0}$, and a small threshold $\epsilon>0$. Apply the initial control policy $u_0 =  - {\mathcal{K}_0}{\eta}+\xi$ with exploration noise $\xi$. Set $j=0$. \\
\textbf{Data Collection.} Collect the sampled data $(u(k),y(k))$ and calculate $(\eta(k), \eta(k+1))$ until the rank condition \eqref{rank condition} holds. \\
\textbf{loop:} \\
\qquad Calculate $\mathcal{P}_{\eta,\eta}^{(j)}$, $\mathcal{P}_{\eta,u}^{(j)}$, and $\mathcal{P}_{u,u}^{(j)}$ by \eqref{VI ILE matrix} \\
\qquad Update $\mathcal{P}_{j+1}$ by \eqref{P_VI_model_free} \\
\qquad \textbf{if } $ \| \mathcal{P}_{j+1} - \mathcal{P}_{j}\| < \epsilon$ \\
\qquad \qquad Calculate $\mathcal{K}_j$ by \eqref{K_VI_model_free} \\
\qquad \qquad \textbf{return} $\tilde{ \mathcal{K}}^* = \mathcal{K}_j$ \\
\qquad \textbf{end if } \\
\qquad     $j = j+1 $ \\
\textbf{end loop}\\
\bottomrule
\end{tabular}
\end{table}

\subsubsection{Model-Free PI Scheme}

Given an initial admissible control gain $\mathcal{K}_0$, for every iteration $j=0,1,2, \dots$, one can obtain
\begin{flalign}\label{PI_model_free_derivation}
& ~ \eta ^\T(k+1) \mathcal{P}_j \eta(k+1) - \eta^\T(k) \mathcal{P}_j \eta(k) \nonumber \\
=&~ 2 \eta^\T(k) M^\T (A-BK_j)^\T P_{j} B \left( u(k)+\mathcal{K}_j \eta (k) \right)  \nonumber \\
& +  \left( u(k)+\mathcal{K}_j \eta (k) \right)^\T B^\T P_{j} B \left( u(k)+\mathcal{K}_j \eta (k) \right) \nonumber \\
&- y^\T(k) Q_y y(k) - \eta^\T(k) \mathcal{K}_{j}^\T R \mathcal{K}_{j} \eta(k).
\end{flalign}
Define $\hat u_j(k)=u(k)+\mathcal{K}_j \eta (k)$. Substituting the data $\eta(k)$ and $\hat u_j(k)$ into the functions defined in \eqref{operations}, according to \eqref{PI_model_free_derivation}, the vector-formed PI-based learning equation is given as
\begin{equation} \label{PI ILE vector}
	{\hat {\varphi}}_j (k) \left[\begin{matrix}
									\mbox{vech}(\mathcal{P}_{j})\\
									\mbox{vec}({\mathcal{P}}_{A,B}^{(j)}) \\
									\mbox{vech}({\mathcal{P}}_{B,B}^{(j)})
									\end{matrix} \right] = \phi (k)  \left[\begin{matrix}
									\mbox{vech}(Q_y)\\
									\mbox{vech}(\left( \mathcal{K}_{j}^\T R \mathcal{K}_{j}\right)
									\end{matrix} \right]
\end{equation}
with 
\begin{flalign*}
    {\hat {\varphi}}_j (k) =& \left[ -\bar \delta _{\eta} (k),~ {2{\delta _{{\eta},{R \hat u_j}}}(k)},~ {\delta _{\hat u_j}}(k)\right] \\
    \phi (k) =& \left[ {\delta _{y}}(k),~ {\delta _{\eta}(k)} \right]\\
    \mathcal{P}_{A,B}^{(j)} =& M^\T (A-BK_j)^\T P_{j} B \\
    \mathcal{P}_{B,B}^{(j)} =& B^\T P_{j} B.
\end{flalign*}

By using the least squares method, the matrix-formed data-based learning equation with the PI scheme is given as
\begin{flalign}\label{PI ILE matrix}
{\hat {\Pi}}_j (k_N) \left[\begin{matrix}
									\mbox{vech}(\mathcal{P}_{j})\\
									\mbox{vec}({\mathcal{P}}_{A,B}^{(j)}) \\
									\mbox{vech}({\mathcal{P}}_{B,B}^{(j)})
									\end{matrix} \right] = \Phi (k_N)  \left[\begin{matrix}
									\mbox{vech}(Q_y)\\
									\mbox{vech}(\left( \mathcal{K}_{j}^\T R \mathcal{K}_{j}\right)
									\end{matrix} \right]
\end{flalign}\noindent
where ${\hat {\Pi}}_j (k_N) = \col \left( {\hat {\varphi}}_j(0), {\hat {\varphi}}_j(1), \dots, {\hat {\varphi}}_j(k_N) \right) \in \mathbb{R}^{N \times n _{\varphi}}$ and $\Phi (k_N)  = \col \left( \phi(0), \phi(1),\dots, \phi (k_N) \right) \in \mathbb{R}^{N \times n _{\phi}}$ with dimension $n _{\phi}=\frac{(n_z+p)(n_z+p+1)}{2}$. 

Combining the model-based PI scheme with the solution to \eqref{PI ILE matrix}, $\mathcal{P}_j$ can be estimated by calculating \eqref{PI ILE matrix},
and the control policy can be updated by
\begin{flalign}\label{K_PI_model_free}
	\mathcal{K}_{j+1} = \left( R + \mathcal{P}_{B,B}^{(j)} \right)^{-1} \left( \left(\mathcal{P}_{A,B}^{(j)}\right)^\T + \mathcal{P}_{B,B}^{(j)} \mathcal{K}_j \right).
\end{flalign}

\begin{table}[bpt] \small
\begin{tabular} {p{0.95\columnwidth}}
\toprule
\emph{Algorithm 4.} Model-Free PI-Based Learning Control \\
\midrule
\textbf{Initialization.} 
%Give the weight matrices $Q_y=Q_y^T \geq 0$ and $R=R^\T >0$ satisfying $(A,\sqrt{C^\T Q_y C})$ is observable. 
Choose an initial admissible control gain $\mathcal{K}_0$. Select a pair $(\mathcal{G}_1, \mathcal{G}_2)$, an initial internal state $\eta(0)$, and a small threshold $\epsilon>0$. Apply the initial control policy $u_0 =  - {\mathcal{K}_0}{\eta}+\xi$ with exploration noise $\xi$. Set $j=0$. \\
\textbf{Data Collection.} Collect the sampled data $(u(k),y(k))$ and calculate $(\eta(k), \eta(k+1))$ until the rank condition \eqref{rank condition} holds. \\
\textbf{loop:} \\
\qquad Calculate $\mathcal{P}_{j}$, $\mathcal{P}_{A,B}^{(j)}$, and $\mathcal{P}_{B,B}^{(j)}$ by \eqref{PI ILE matrix}. \\
\qquad \textbf{if } $j\geq 1$ \textbf{and} $ \| \mathcal{P}_{j} - \mathcal{P}_{j-1}\| < \epsilon$ \\
\qquad \qquad \textbf{return} $\tilde{ \mathcal{K}}^* = \mathcal{K}_j$ \\
\qquad \textbf{else } \\
\qquad  \qquad Update $\mathcal{K}_{j+1}$ by \eqref{K_PI_model_free}. \\
\qquad \textbf{end if} \\
\qquad  $j = j+1 $ \\
\textbf{end loop}\\
\bottomrule
\end{tabular}
\end{table}

%Under the observability condition, the data information of $(z, z)$ is available by the known user-defined dynamics \eqref{z_all_defined_or} where the original system input $u$ and output $y$ are seen as the input of \eqref{z_all_defined_or}. Thus, by only accessing the measurable data information of input $u$ and output $y$, the iterative learning equation can be established for all $k \geq 0$. Compared to \cite{Rizvi2019, Rizvi2020_aut,   Chen2022, Chen2023} where the iterative learning equation can not be ideally established due to the existence of unknown observer error, the iterative learning equation in this paper is efficient to the proposed learning control method for $k \geq 0$. 

%By solving the iterative learning equation \eqref{LS_implementation}, the estimated control gain $\bar K_j = K_j M$ for the dynamic output feedback controller \eqref{u_form_MF} is obtained. This implies that not only the control gain $K_j$ but also the unknown constant matrices in \eqref{virtual observer error transformation} and \eqref{x_v_to_z} for the observed state $\hat x$ and the observer error $\varepsilon_x$ are approximated. 

We next show that the same rank condition \eqref{rank condition} can be used to determine whether the collected data is enough for implementing the learning process of the model-free PI scheme.

\begin{theorem} \label{theorem_rank condition PI}
If there exists an instant $k_{N}^*$ such that the rank condition \eqref{rank condition} is satisfied, then, for every iteration $j = 0, 1, 2, \dots$, the PI-based learning equation \eqref{PI ILE matrix} has a unique solution. 
\end{theorem}

\emph{Proof.} 
%Since The data regressor ${\hat {\Pi}}_j (k_N^*)$ is dependent on the updated control gain $\mathcal{K}_j$, which makes it more difficult to analyze the rank of ${\hat {\Pi}}_j(k_N^*)$ for every iteration $j=0,1, 2, \dots$. In what follows, 
We give a proof by contradiction to show the existence of a unique solution to \eqref{PI ILE matrix} for every iteration.
Suppose that ${\hat {\Pi}}_j(k_N^*)$ in \eqref{PI ILE matrix} is not of full column rank such that \eqref{rank condition} is satisfied. This implies that
\begin{flalign}\label{tp1}
{\hat {\Pi}}_j(k_N^*) \left[\begin{matrix}
									\mbox{vech}(\mathcal{P}_{j})\\
									\mbox{vec}({\mathcal{P}}_{A,B}^{(j)}) \\
									\mbox{vech}({\mathcal{P}}_{B,B}^{(j)})
									\end{matrix} \right] = \mathbf{0}
\end{flalign}\noindent
has a nonzero solution denoted by $\mathbb{W}  = \col \left(V,~W_1,~W_2 \right) $ with $V = \mbox{vech}({M^\T V^m M})$, $W_1 = \mbox{vec}({M^\T W_1^m})$ and $W_2 = \mbox{vech}({W_2^m})$. Substituting the data $\eta(k)$, $u(k)$ and $\hat u_j(k)$ into the data regressor functions defined in \eqref{operations regressor}, 
it follows from \eqref{tp1} that
\begin{flalign}\label{tp2}
{\bar \Delta _{\eta}} V -
2\Gamma_{\eta,\hat u_j} W_1 - \Delta_{\hat u_j} W_2  = \mathbf{0}.
\end{flalign}\noindent

By \eqref{sys_origin_x} and \eqref{transformation x eta}, the first part in the left-hand side of \eqref{tp2} becomes 
\begin{flalign}
	{\bar \Delta _{\eta}}V \!=&  
 \Delta _{\eta} \mbox{vech} \left( M^\T \left( \left( A \!-\! B K_{j} \right)^\T \! V^m \! \left( A \!-\! B K_{j} \right) \!-\! V^m \right) M \right) \nonumber \\
 & +\Delta _{\eta} \mbox{vech}\left( M^\T \left( 2 A ^\T V^m B K_j \!-\! K_j^\T B ^\T V^m B K_j \right)M \right) \nonumber \\
& + 2\Gamma _{\eta, u} \mbox{vec}\left( M^\T ( A ^\T V^m B) \right)
\!+\!   \Delta _{u} \mbox{vech}( B ^\T V^m B). \label{tp3}
\end{flalign}

Combining \eqref{tp2} and \eqref{tp3}, we have
\begin{flalign}\label{tp4}
{\Delta _{{\eta}}}\mbox{vech}(M^\T {\Omega _1} M) 
\!+\! 2{\Gamma _{{\eta},{u}}} \mbox{vec}(M^\T {\Omega _2}) 
\!+\! {\Delta _{{u}}}\mbox{vech}({\Omega _3}) \!=\! \mathbf{0}
\end{flalign}\noindent
where
\begin{subequations}\label{tp5}
\begin{flalign}
    {\Omega _1} =&  \left( A \!-\! B K_{j} \right)^\T \! V^m \! \left( A \!-\! B K_{j} \right) - V^m \nonumber \\
&+ 2\left( A ^\T V^m  B - W_1^m - \left( K_j \right)^\T W_2^m\right) K_j \label{t4_w1} \\
&- \left(  K_j \right)^\T \left( B ^\T V^m  B - W_2^m \right) K_j  \nonumber \\
\label{t4_w2} {\Omega _2} =& A ^\T V^m  B - W_1^m - \left( K_j \right)^\T W_3^m  \\
\label{t4_w3} {\Omega _3} =& B ^\T V^m B - W_2^m.
\end{flalign} 
\end{subequations}
If the rank condition \eqref{rank condition} is satisfied, then \eqref{tp4} only has zero solution, i.e., ${\Omega _1}=\mathbf{0}$, ${\Omega _2}=\mathbf{0}$ and ${\Omega _3}=\mathbf{0}$. Then, we have
\begin{flalign}\label{tp8}
\begin{array}{l}
\left( A \!-\! B K_{j} \right)^\T \! V^m \! \left( A \!-\! B K_{j} \right) - V^m = \mathbf{0}.
\end{array}
\end{flalign}\noindent
According to \cite{Hewer1971} that the matrix $\left( A - B K_{j} \right)$ is Schur for all $j =0, 1, 2, \dots$, the only solution to \eqref{tp8} is ${V^m} = \mathbf{0}$. Substituting $V^m = \mathbf{0}$ into \eqref{tp5}, we have ${W_1^m} = \mathbf{0}$, ${W_2^m} = \mathbf{0}$, and ${W_3^m} = \mathbf{0}$, which contradicts the fact that $\mathbb{W}$ is a nonzero solution to \eqref{tp1}. Thus, if \eqref{rank condition} holds, then, for all $j =0, 1, 2, \dots$, ${\hat{\Pi}}_{j} (k_N^*)$ is of full rank. \hfill \rule{1.5mm}{1.5mm}

%Thus, there always exists a unique solution to both the VI-based and the PI-based learning equations \eqref{VI ILE matrix} and \eqref{PI ILE matrix} when the rank condition \eqref{rank condition} is met.  This completes the proof. \hfill \rule{1.5mm}{1.5mm}

Finally, the proposed output feedback learning control approach with the PI scheme is summarized in Algorithm 4.

\section{Convergence, Stability, and Optimality Analysis}
In this section, we give a detailed analysis of convergence, stability, and optimality of the proposed output feedback learning control approaches and provide a model-free stability criterion. Moreover, we propose a switched iteration scheme integrated with PI and VI 
by leveraging the stability criterion, which can relax
the requirement of an initial admissible controller in the PI scheme and accelerate the convergence rate in the VI scheme.

\subsection{Convergence analysis}
To show the convergence of the learning algorithm, we first give a rank condition for the parameterization matrix $M$ as follows:
\begin{assumption} \label{ass rank condition M }
The parameterization matrix $M$ satisfies
\begin{flalign}\label{rank condition M}
    \textnormal{rank} (M) = n.
\end{flalign}
\end{assumption}

\begin{remark}
In {\rm \cite{Rizvi2023TAC}}, a necessary condition for the convergence of learning algorithms with both PI and VI is given. That is, the parameterization matrix $\bar W$ in \eqref{state parameterization} is of full rank.
It is noted that the state observer error parameterization matrix $M_{\varepsilon_x \leftarrow \eta_\varepsilon}$ is constructed in $M$.
Thus, compared with the full rank condition of $\bar W$, condition \eqref{rank condition M} is easier to hold.
 \end{remark}

We next show the full rank \eqref{rank condition M} in Lemma \ref{lemma rank full for M_error} can be guaranteed by providing a simple design method.  

\begin{lemma} \label{lemma rank full for M_error}
If the matrix $A_L$ has $n$ distinct eigenvalues, then there exists a square matrix $M_{\varepsilon_x \leftarrow \eta_\varepsilon}$ that is of full rank. 
\end{lemma}

\emph{Proof.} It follows from \eqref{virtual observer error transformation} that the square matrix $M_{\varepsilon_x \leftarrow \eta_\varepsilon} = M_{\varepsilon_x} M_{\eta_{\varepsilon}}^{-1}$ is of full rank if and only if $M_{\varepsilon_x}$ is non-singular, since $M_{\eta_{\varepsilon}}$ is a user-defined non-singular matrix. Let $\varepsilon_0 = \varepsilon_x (0)$ be the initial observer error. Then, we give the existence proof of $M_{\varepsilon_x}$ being of full rank as follows.

The determinant and adjoint matrix of $(zI_n-A_L)$ can be rewritten as
\begin{flalign}
\label{det}    \mbox{det}(zI_n\! -\! A_L) &\! =\!  z^{n}\! +\! b_{1} z^{n\! -\! 1} \!+\! b_{2} z^{n\! -\! 2} \!+\! \cdots \!+\! b_{n-1}z \!+\! b_n \\
\label{adj}     \mbox{adj}(zI_n\! -\! A_L) &\! =\!  \Upsilon_{1} z^{n\! -\! 1}+\Upsilon_{2} z^{n\! -\! 2} + \cdots + \Upsilon_{n\! -\! 1} z \! +\!  \Upsilon_{n}
\end{flalign}
where $b_i\in \mathbb{R}$ and $\Upsilon_i\in \mathbb{R}^n$, $i=1,2,\dots,{n}$, are the constant coefficients and matrix coefficients, respectively. 
According to the properties of the determinant and adjoint matrix, one can obtain
\begin{flalign}\label{equality_det adj} 
    &\mbox{det}(zI_n-A_L) I_n \nonumber \\
    =& \mbox{adj}(zI_n-A_L) (zI_n-A_L) \nonumber \\
    =& \Upsilon_1 z^{n}+ (\Upsilon_2 - \Upsilon_{1}A_L) z^{n-1} +\cdots + (\Upsilon_{i+1} - \Upsilon_{i}A_L) z^{n-i} \nonumber \\
    &+\cdots+(\Upsilon_{n} - \Upsilon_{n-1}A_L) z -  \Upsilon_{n}A_L.
\end{flalign}

From the equality of the coefficients of \eqref{det}--\eqref{equality_det adj}, we have
\begin{flalign*}
    \Upsilon_{i+1} = \Upsilon_{i}A_L + b_i I_n, ~~~~i=1,2,\dots, n-1
\end{flalign*}
with $\Upsilon_1 = I_n$.

Then, $M_{\varepsilon_x}$ can be represented by
\begin{flalign*}
    M_{\varepsilon_x}
    =& \left[ \Upsilon_{n} \varepsilon_{0}, ~ \Upsilon_{n-1} \varepsilon_{0}, ~\dots~, ~\Upsilon_{3} \varepsilon_{0}, ~\Upsilon_{2} \varepsilon_{0},~ \Upsilon_{1} \varepsilon_{0}, \right] \nonumber \\
    =& \left[A_L^{n-1} \varepsilon_0+ b_1 A_L^{n-2}\varepsilon_0+ \cdots + b_{n-1} \varepsilon_0, \right. \nonumber \\
    &~\left. ~\dots, A_L^2 \varepsilon_{0} + b_1 A_L\varepsilon_{0}+ b_2 \varepsilon_{0}, ~A_L \varepsilon_{0} + b_1 \varepsilon_{0}, ~ \varepsilon_{0} \right].
\end{flalign*}
Thus, we have
\begin{flalign*}
\mbox{rank} \left( M_{\varepsilon_x} \right) 
    = \mbox{rank}\left( \left[ \varepsilon_{0}, ~A_L \varepsilon_{0}, ~    A_L^2 \varepsilon_{0}, ~ \dots,~ A_L^{n-1} \varepsilon_{0} \right] \right). 
\end{flalign*}
If the matrix $A_L$ has $n$ distinct eigenvalues, i.e., $\lambda_1, \lambda_2, \dots, \lambda_n$ are not equal to each other, then there exists a non-singular matrix $U$ which can transform $A_L$ into a diagram matrix $\mathcal{J}_L = \mbox{block diag} [\lambda_1, \lambda_2, \dots, \lambda_n]$, i.e., $U^{-1}A_L U = \mathcal{J}_L$. 
%It is clear that $ \mathcal{J}_L ^ i = \left( U^{-1}A_L U \right)^i = U^{-1}A_L^i U$.
Letting $\bar \varepsilon_0 = U^{-1} \varepsilon_{0}$, we have
\begin{flalign*}
    U^{-1} &\left[ \varepsilon_{0}, ~A_L \varepsilon_{0}, ~    A_L^2 \varepsilon_{0}, ~ \cdots,~ A_L^{n-1} \varepsilon_{0} \right] \nonumber \\
    =&\left[ \bar \varepsilon_0, ~ U^{-1} A_L U \bar \varepsilon_0, ~    U^{-1} A_L^2  U \bar \varepsilon_0, ~ \cdots,~ U^{-1} A_L^{n-1}  U \bar \varepsilon_0 \right] \nonumber \\
    =&\left[ \bar \varepsilon_0, ~ \mathcal{J}_L \bar \varepsilon_0, ~    \mathcal{J}_L ^ 2 \bar \varepsilon_0, ~ \cdots,~ \mathcal{J}_L^{n-1} \bar \varepsilon_0 \right] \nonumber \\
    =&\mbox{dia} (\bar \varepsilon_0) \left[
\begin{matrix}
     1 &  \lambda_1 & \lambda_1^2 & \cdots &  \lambda_1^{n-1} \\
     1 &  \lambda_2 & \lambda_2^2 & \cdots &  \lambda_2^{n-1} \\
     \vdots &  \vdots & \vdots & \cdots &  \vdots \\
     1 &  \lambda_n & \lambda_n^2 & \cdots &  \lambda_n^{n-1} \\
\end{matrix} \right].
\end{flalign*}
If all the elements of $\bar \varepsilon_0$ are nonzero, the matrix $M_{\varepsilon_x}$ is of full rank. Therefore, since $M_{\eta_{\varepsilon}}$ is chosen as non-singular, there exists a matrix $M_{\varepsilon_x \leftarrow \eta_\varepsilon} =M_{\varepsilon_x} M_{\eta_{\varepsilon}}^{-1} $ that is of full rank. This completes the proof.  \hfill \rule{1.5mm}{1.5mm}

\begin{remark}
In some existing output feedback learning algorithms, for instance, {\rm \cite{Rizvi2023TAC, Rizvi2020_aut, Chen2023, Gao2016-3aut}}, the constructed parameterization matrices are in the form of either $\bar M$ in \eqref{state reconstruction} or $\bar W$ in \eqref{state parameterization}. Under Assumption \ref{ass_1}, it follows from \eqref{state reconstruction} that the matrix $\bar M$ may not be of full row rank when the system matrix $A$ has the eigenvalue $0$, and the matrix $\bar W$ may not be of full row rank when the observer matrices $(A - LC)$ and $A$ have common eigenvalues according to {\rm \citet[Theorem 4]{Rizvi2023TAC}}. Different from the parameterization matrices $\bar M$ and $\bar W$, 
the rank condition of parameterization matrix $M$ in this paper can be satisfied by choosing the constructed $M_{\varepsilon_x \leftarrow z_\varepsilon}$ to be full rank. According to Lemma \ref{lemma rank full for M_error}, there exists matrix $M_{\varepsilon_x \leftarrow z_\varepsilon}$ that has full rank by setting all the eigenvalues of $A_L$ distinct. This pole assignment can be easily achieved which only requires Assumption \ref{ass_1} satisfied.
\end{remark}

\begin{remark}
The exploration noise is usually added to the control input during the data collection process to ensure the rank condition \eqref{rank condition}, for instance, {\rm \cite{Bian2016,Jiang2012,Modares2016,Rizvi2023TAC}}. Thus, it is not difficult to get a nonzero state $U^{-1}{\varepsilon}_x(k)$ at an instant $k$ that can be set as the initial instant $0$ for the learning process. Furthermore, we provide a way to check if the rank condition \eqref{rank condition M} of $M$  is met or not in Subsection \ref{sec stability analysis}.
%Moreover, the non-singularity of $M_{\varepsilon_x \leftarrow \eta_\varepsilon}$ can be easily verified in the learning process by checking $ \mbox{det}(\mathcal{H}_j(K_j, P_j)) \neq 0$. 
%. Specifically, by computing \eqref{LS_implementation}, the matrix $\bar H_j$ is obtained. Note that $\tilde H_j = M_{\varepsilon_x \leftarrow z_\varepsilon}^\T (A^\T P_j + P_j A + Q_x) M_{\varepsilon_x \leftarrow z_\varepsilon}$ is a block matrix of $\bar H_j$. By checking $\mbox{det}( \tilde H_j ) \neq 0$, the non-singularity information of the square matrix $M_{\varepsilon_x \leftarrow z_\varepsilon}^\T$ is available. 
\end{remark}

Now, we are ready to give the convergence analysis of Algorithms 3 and 4 shown in the following Theorems \ref{theorem convergence analysis VI} and \ref{theorem convergence analysis PI}, respectively. 

\begin{theorem} \label{theorem convergence analysis VI}
For any given initial positive semi-definite symmetric matrix $\mathcal{P}_0$, if the rank conditions \eqref{rank condition} and \eqref{rank condition M} hold, the control gain $\mathcal{ K}_j$ estimated by the model-free VI-based Algorithm 3 converges to the optimal value $\mathcal{K}^*$, i.e., $\lim_{j \rightarrow \infty} \mathcal{K}_j= \mathcal{K}^*$.
\end{theorem}

\emph{Proof.} By the solution
to the iterative learning equation \eqref{VI ILE matrix}, the value evaluation equation \eqref{P_VI_model_free} can be written as
\begin{flalign*}
\mathcal{P}_{j+1} =  M^\T \left( A^\T  P_{j}  A + C^\T Q_y C - A^\T  P_j  B \left( R + B^\T  P_j B \right)^{-1} B^\T  P_j A \right) M 
\end{flalign*}
If the rank condition \eqref{rank condition M} holds, \eqref{P_VI_model_free} is equivalent to \eqref{VI_model_based}.
As proven in \cite{dorato1971optimal}, for any given $P_0= P_0^\T \geq 0$, the $P_j$ calculated by \eqref{VI_model_based} converges to $\bar P^*$ as $j \rightarrow \infty$. In Algorithm 3, the value evaluation equation \eqref{P_VI_model_free} with the solution to the iterative learning equation \eqref{VI ILE matrix} is transformed from \eqref{VI_model_based}. As shown in Theorem \ref{theorem_rank condition VI}, the existence of the unique solution of \eqref{VI ILE matrix} can be ensured once the rank condition \eqref{rank condition} is satisfied. Thus, in Algorithm 3, for any given ${\mathcal{P}_0= \mathcal{P}_0^\T \geq 0}$, substituting ${ \mathcal{P}_j}$ into \eqref{VI ILE matrix} for each iteration $j = 0,1, \dots $, ${\mathcal{P}_{j + 1}}$ can be recalculated uniquely from \eqref{P_VI_model_free}. Therefore, as $j \rightarrow \infty$, $ \mathcal{P}_{j+1}$ converges to ${ \mathcal{P}^ * }$, which implies that $ \mathcal{K}_j$ converges to $\mathcal{K}^*$.
Thus, the proof is completed.	  \hfill \rule{1.5mm}{1.5mm}

\begin{theorem} \label{theorem convergence analysis PI}
For any given initial admissible control policy $\mathcal{K}_0$, if the rank conditions \eqref{rank condition} and \eqref{rank condition M} hold, the control gain $\mathcal{ K}_j$ estimated by the model-free PI-based Algorithm 4 converges to the optimal value $\mathcal{K}^*$, i.e., $\lim_{j \rightarrow \infty} \mathcal{K}_j= \mathcal{K}^*$.
\end{theorem}

\emph{Proof.} Along the similar way as the proof in Theorem \ref{theorem convergence analysis VI}, when the rank conditions \eqref{rank condition} and \eqref{rank condition M} are satisfied, it is not hard to have $\lim_{j \rightarrow \infty} \mathcal{K}_j= \mathcal{K}^*$ with Algorithm 4 by the convergence of the PI scheme proven in \cite{Hewer1971}. Thus, the proof is omitted.  \hfill \rule{1.5mm}{1.5mm}

%\begin{remark}
%According to Theorems \ref{theorem convergence analysis VI} and \ref{theorem convergence analysis PI}, the convergence of Algorithms 3 and 4 only requires the rank condition to be satisfied. The estimated optimal control policy converges to its optimal value, which achieves the stability of the closed-loop system. It should be pointed out that the matrix $M_{\varepsilon_x \leftarrow \eta_\varepsilon}$ does not affect the stability even if $M_{\varepsilon_x \leftarrow \eta_\varepsilon}$ is singular. The requirement on the non-singularity of the matrix $M_{\varepsilon_x \leftarrow \eta_\varepsilon}$ is only an auxiliary condition for the stability criterion \eqref{stability condition} of the output feedback learning control.
%\end{remark}

\begin{remark}
Different from the parameterization matrix constructed in {\rm \cite{Rizvi2023TAC, Chen2022}}, we develop a method to construct a square matrix $M_{\varepsilon_x \leftarrow \eta_\varepsilon}$ that is of full rank. It can not only increase the flexibility of the matrix $M$ but also enable us to analyze the stability of the closed-loop system with the estimated control policy at any iteration $j$ shown in Subsection \ref{sec stability analysis}. 
\end{remark}

\subsection{Stability Analysis} \label{sec stability analysis}

By using a stopping criterion with a small threshold $\epsilon$, we can obtain an estimated control gain $\tilde {\mathcal{K}}^* = {\tilde K}^* M$ by Algorithm 3 or Algorithm 4. For the PI-based scheme, as proven in \cite{Hewer1971}, the estimated control policy has the property that $\sigma(A-B{\tilde K}^*) \in \mathbb{U}^-$ for $j=0,1,2,\dots$. This indicates that the stability of the closed-loop system with the PI-based scheme can be guaranteed at every iteration $j$. 

However, for the VI-based scheme, this guaranteed stability property does not hold. Moreover, it is important to notice that the matrix $\left( C^\T Q_y C + (K^*)^\T R K^* \right)$ is positive semi-definite in some cases, for example, the dimension of the system satisfies $n>p+m$. In these cases, the estimated matrix $\left( \left(A-B{\tilde K}^* \right) ^\T \hat P^* \left(A-B{\tilde K}^* \right)-\tilde P^* \right)$ by using the stopping criterion may not be positive semi-definite. This implies that the closed-loop system with the estimated control gain $\tilde {\mathcal{K}}^*$ by Algorithm 3 may not achieve globally asymptotically stable at the origin. Thus, it is worth providing a stability criterion for the closed-loop system with the proposed output feedback learning control approach.

To analyze the stability of the linear discrete-time closed-loop systems, we recall the classical Lyapunov theorem.

\begin{lemma} \cite{chen1984linear} \label{theorem lyapunov}
Given any positive definite symmetric matrix $Q\in \mathbb{R}^{n \times n}$, there exists a unique positive definite symmetric solution $P$ to the Lyapunov equation  
\begin{flalign*}
    \left( A - B K \right)^\T  P  \left( A - B K \right) - P = -Q
\end{flalign*}
if and only if $\sigma(A - B K)\in \mathbb{U}^-$.
\end{lemma}

In the case of completely unknown system matrices, an alternative way to analyze the stability of the closed-loop systems is to construct an equivalent Lyapunov equation by using the data information. Although the data of states, as the most important information that can reflect system dynamics, is unmeasurable, it is fortunate that there always exists a linear relationship between the designed internal state $\eta$ and the real state $x$ as shown in \eqref{transformation x eta}. 
By using the solution to an auxiliary data-based equation, we next provide a model-free stability criterion for the closed-loop system under the estimated control policy at any iteration $j$.

Define the auxiliary data-based equation as
\begin{equation}\label{stability ILE}
	\varphi (k) \left[\begin{matrix}
									\mbox{vech} (\bar {\mathcal{P}}_{\eta,\eta}^{(j)})\\
									\mbox{vec}({\mathcal{P}}_{\eta,u}^{(j)}) \\
									\mbox{vech}({\mathcal{P}}_{u,u}^{(j)})
									\end{matrix} \right] = \psi (k) \mbox{vech} \left( {\mathcal{P}}_j \right)+ \bar {\mathcal{C}}(k)
\end{equation}
with $ \bar {\mathcal{P}}_{\eta,\eta}^{(j)} = \mathcal{P}_{\eta,\eta}^{(j)} = M^\T A^\T  P_{j} A M$ and $\bar {\mathcal{C}}(k) = u^\T(k) R u(k)$, which is the modification of \eqref{VI ILE vector} with utility function $\bar {\mathcal{C}}(k)$. By the solution to \eqref{stability ILE}, we define 
\begin{flalign*}
{H}_j ( P_j, K_j) = \bar {\mathcal{P}}_{\eta,\eta}^{(j)} - 2 {\mathcal{P}}_{\eta,u}^{(j)}  \mathcal{K}_j \! + \! \left( \mathcal{K}_j \right) ^\T {\mathcal{P}}_{u,u}^{(j)}  \mathcal{K}_j - \mathcal{P}_j
\end{flalign*}
where $\mathcal{K}_{j} = \left( \mathcal{P}_{u,u}^{(j)} \right)^{-1} \left(\mathcal{P}_{\eta,u}^{(j)}\right)^\T$.
Then, by extracting the last $n \times n$ submatrix from ${H}_j ( P_j, K_j)$ to form a new matrix $\mathcal{H}_j ( P_j, K_j)$, one can obtain
\begin{flalign*}
    \mathcal{H}_j ( P_j, K_j) 
\!= M_{\varepsilon_x \leftarrow \eta_\varepsilon}^\T \left( \left( A \!-\! B K_j \right)^\T \! P_j \! \left( A \!-\! B K_j \right) \!-\! P_j \right) M_{\varepsilon_x \leftarrow \eta_\varepsilon}.
\end{flalign*}
Note that the non-singularity of $M_{\varepsilon_x \leftarrow \eta_\varepsilon}$ can be easily verified in the learning process by checking $ \mbox{det}(\mathcal{H}_j(K_j, P_j)) \neq 0$.

Then, we provide a way to guarantee the stability of the closed-loop system with the estimated control policy by VI-based model-free algorithm. 

%As we stated before, although an estimated optimal control gain $\tilde {\mathcal{K}}^*$ can be obtained by using a stopping criterion with a small threshold $\epsilon$ in Algorithm 3, the stability criterion may not be satisfied in the case of positive semi-definite matrix $(Q_x + (K^*)^\T R K^*)$. 
%In what follows, we provide a way to guarantee the stability of the closed-loop system with the estimated control policy by Algorithm 3.

\begin{lemma} \label{corollary modified C}
Given any symmetric positive definite matrix $Q_\epsilon$, under rank condition \eqref{rank condition} and \eqref{rank condition M}, then there always exists an iteration $j$ satisfying the stability criterion
\begin{flalign} \label{stability condition}
                \mathcal{H}_j ( P_j, K_j) < 0
\end{flalign}\noindent
by employing the model free VI-based scheme with the utility function 
\begin{flalign}\label{cost_function modify 1}
    \tilde {\mathcal{C}}(k) =  \eta ^\T(k) Q_\epsilon \eta (k) + y^\T(k) Q_y y(k) +  u^\T(k) R u(k).
\end{flalign}
\end{lemma}

\emph{Proof.} We give a proof by contradiction for the existence of the iteration $j$. Let $\bar Q_x = (M M^\T)^{-1} M Q_\varepsilon M^\T (M M^\T)^{-1}$ and ${\bar Q}_\varepsilon = Q_\varepsilon - M^\T {\bar Q}_x M$. The utility function \eqref{cost_function modify 1} can be rewritten as
\begin{flalign*}
    \tilde {\mathcal{C}}(k) =  \eta^\T (k) {\bar Q}_\epsilon \eta (k) + x^\T(k) {\tilde Q}_x x(k) +  u^\T(k) R u(k)
\end{flalign*}
with ${\tilde Q}_x = {\bar Q}_x+ C^\T Q_y C$. Under the rank condition \eqref{rank condition}, according to Theorem \ref{theorem_rank condition VI}, the iterative learning equation \eqref{VI ILE matrix} with the utility function \eqref{cost_function modify 1} has a unique solution at any iteration $j$. With the solution to \eqref{VI ILE matrix},
the value evaluation \eqref{P_VI_model_free} can be rewritten as
%Let's see the matrix $\mathcal{Q} = Q_\epsilon + M^\T C^\T Q_y C M$ as the weight matrix, then 
%The value evaluation \eqref{P_VI_model_free} can be rewritten as
\begin{flalign} \label{corollary 1 eq1}
 \mathcal{P}_{j+1}
 = M^\T \! \left( A^\T  P_{j} A \!+\! {\tilde Q}_x \!-\! A^\T  P_{j} B \left(R \!+\! B^\T  P_{j} B \right)^{-1} \! B^\T  P_{j} A \right) \! M \!+\! {\bar Q}_\varepsilon.
\end{flalign}
Under the rank condition \eqref{rank condition M}, i.e., $M$ is of full row rank, it follows from \eqref{corollary 1 eq1} that
\begin{flalign}\label{P_VI_model_based col 1}
 &(M M^\T)^{-1} M \mathcal{P}_{j+1} M^\T (M M^\T)^{-1}  \nonumber \\
=&A^\T  P_{j} A \!+\! {\tilde Q}_x \!-\! A^\T  P_{j} B \left(R \!+\! B^\T  P_{j} B \right)^{-1} B^\T  P_{j} A 
:= P_{j+1}.
\end{flalign}
The control policy update law at iteration $j$ is given as
\begin{flalign}\label{K_VI_model_based col 1}
	{K}_{j} = \left( R \!+\! B^\T  P_{j} B \right)^{-1} B^\T  P_{j} A.
\end{flalign}
%Note that the value evaluation \eqref{P_VI_model_based} and control update law \eqref{K_VI_model_based} form a classic VI scheme shown in \cite{Lancaster1995} where $\lim_{j \rightarrow \infty} P_j = P^*$ is guaranteed for any initial positive definite symmetric matrix $P^0$. 

Substituting \eqref{K_VI_model_based col 1} into \eqref{P_VI_model_based col 1}, one can obtain
\begin{flalign*}
 {P}_{j+1} -P_j  
 = \left( A \!-\! BK_j\right) ^\T  P_{j} \left( A \!-\! BK_j\right) \!-\! P_j \!+\! {\tilde Q}_x \!+\! K_j ^\T R  K_{j}.
\end{flalign*}
According to \rm{\cite{dorato1971optimal}} that $\lim_{j \rightarrow \infty} P_j = P^*$, we have 
\begin{flalign} \label{P_VI_model_based_theorem4 3}
\lim_{j \rightarrow \infty} \! \left( A \!-\! BK_j\right) ^\T \! P_{j} \left( A \!-\! BK_j\right) \!-\! P_j \!= \! - {\tilde Q}_x \!-\! K_j ^\T R  K_{j}. 
\end{flalign}
Note that ${\tilde Q}_x>0$ since $Q_\varepsilon >0$ and $Q_y \geq 0$. If \eqref{stability condition} is not satisfied for any iteration $j$, it contradicts the fact shown in \eqref{P_VI_model_based_theorem4 3} where ${\tilde Q}_x$ is positive definite.  This completes the proof.  \hfill \rule{1.5mm}{1.5mm}

Finally, by Lemmas \ref{theorem lyapunov} and \ref{corollary modified C}, we are ready to show the stability of the system as follows.

\begin{theorem} \label{theorem stability} 
Under the rank condition \eqref{rank condition} and \eqref{rank condition M}, there exists  an iteration $j$ such that the closed-loop system under the output feedback control $u(k)=-\mathcal{K}_j \eta(k)$ is asymptotically stable.
\end{theorem}

\emph{Proof.} Under the rank condition \eqref{rank condition} and \eqref{rank condition M}, by Lemma \ref{corollary modified C}, 
for any symmetric positive definite matrix $Q_\epsilon$,
there always exists an
iteration such that 
% If the inequality 
\eqref{stability condition} holds.
Then, there exists a positive definite symmetric matrix $\mathcal{Q}$ satisfying
\begin{flalign*}
    M_{\varepsilon_x \leftarrow \eta_\varepsilon}^\T \left( \left( A \!-\! B K_j \right)^\T \! P_j \! \left( A \!-\! B K_j \right) \!-\! P_j \right) M_{\varepsilon_x \leftarrow \eta_\varepsilon} = - \mathcal{Q}.
\end{flalign*}
By Lemma \ref{theorem lyapunov}, we have $\sigma(A \!-\! B K_j) \in \mathbb{U}^-$. Combining Theorem \ref{theorem control design}, the closed-loop system under the output feedback control $u(k)=-\mathcal{K}_j \eta(k)$ is asymptotically stable. This completes the proof. \hfill \rule{1.5mm}{1.5mm}

Furthermore, combining Lemma \ref{corollary modified C} with the model-free VI scheme and model-free PI scheme, we propose an output feedback learning control approach with a model-free switched iteration (SI) scheme in Algorithm 5.

\begin{remark}
Compared to the VI scheme, the PI scheme has a faster convergence rate while the initial admissible control policy is required. The proposed model-free SI scheme provides a way to combine the VI scheme with the PI scheme, which takes advantage of both PI and VI schemes to some extent while avoiding their disadvantages.
\end{remark}

%\begin{remark}
%The non-singularity of $M_{\varepsilon_x \leftarrow \eta_\varepsilon}$ can be easily verified in the learning process by checking $ \mbox{det}(\mathcal{H}_j(K_j, P_j)) \neq 0$. 
%Specifically, by computing \eqref{stability ILE}, the matrix $\bar H_j$ is obtained. Note that $\tilde H_j = M_{\varepsilon_x \leftarrow z_\varepsilon}^\T (A^\T P_j + P_j A + Q_x) M_{\varepsilon_x \leftarrow z_\varepsilon}$ is a block matrix of $\bar H_j$. By checking $\mbox{det}( \tilde H_j ) \neq 0$, the non-singularity information of the square matrix $M_{\varepsilon_x \leftarrow z_\varepsilon}^\T$ is available. 
%\end{remark}

\begin{table}[bpt] \small
\begin{tabular} {p{0.95\columnwidth}}
\toprule
\emph{Algorithm 5.} Model-Free SI-Based Learning Control \\
\midrule
\textbf{Initialization.}  %Give the weight matrices $Q_y=Q_y^T \geq 0$, $Q_\varepsilon = Q_\varepsilon^\T >0$ and $R=R^\T >0$ satisfying $(A,\sqrt{C^\T Q_y C})$ is observable. 
Define a utility function as $\mathcal{C}(k)$ in \eqref{cost_function} and a modified utility function as $\tilde {\mathcal{C}}(k)$ in \eqref{cost_function modify 1} with $Q_\varepsilon = Q_\varepsilon^\T >0$. Select a pair $(\mathcal{G}_1, \mathcal{G}_2)$, an initial internal state $\eta(0)$, an initial control gain $\mathcal{K}_0$, and a small threshold $\epsilon>0$. Apply the initial control policy $u_0 =  - {\mathcal{K}_0}{\eta}+\xi$ with exploration noise $\xi$. Set $\textbf{sign}_{\rm{PI}} = 0$ and $j=0$. \\
\textbf{Data Collection.} Collect the sampled data $(u(k),y(k))$ and calculate $(\eta(k), \eta(k+1))$ until the rank condition \eqref{rank condition} holds. \\
\textbf{loop:} \\
\qquad \textbf{if} $\textbf{sign}_{\rm{PI}} = 1$ \\
\qquad \qquad Calculate $\mathcal{P}_{j}$, $\mathcal{P}_{A,B}^{(j)}$, and $\mathcal{P}_{B,B}^{(j)}$ by \eqref{PI ILE matrix} with $\mathcal{C}(k)$. \\
\qquad \qquad Update $\mathcal{K}_{j+1}$ by \eqref{K_PI_model_free}. \\
\qquad \textbf{else} \\
\qquad \qquad Calculate $\mathcal{P}_{\eta,\eta}^{(j)}$, $\mathcal{P}_{\eta,u}^{(j)}$, and $\mathcal{P}_{u,u}^{(j)}$ by \eqref{VI ILE matrix} with $\tilde {\mathcal{C}}(k)$. \\
\qquad \qquad Update $\mathcal{K}_j$ and $\mathcal{P}_{j+1}$ by \eqref{K_VI_model_free}
	and \eqref{P_VI_model_free}, respectively. \\
\qquad \qquad \textbf{if} $\mathcal{H}_j (P_j, K_j) < 0$ \\
\qquad \qquad \qquad $\textbf{sign}_{\rm{PI}} = 1$, ~~ \textbf{continue} \\
\qquad \qquad \textbf{end if} \\
\qquad \textbf{end if }\\
\qquad \textbf{if } $j \geq 1 $ \textbf{and} $ \| \mathcal{P}_{j} - \mathcal{P}_{j-1}\| < \epsilon$ \\
\qquad \qquad \textbf{return} $\tilde {\mathcal{K}}^* = \mathcal{K}_j$ \\
\qquad \textbf{end if} \\
\qquad \qquad     $j = j+1 $ \\
\textbf{end loop}\\
\bottomrule
\end{tabular}
\end{table}

\subsection{Optimality Analysis}

To analyze the optimality of several output feedback controllers with respect to the pre-defined value function \eqref{cost_function}, the optimality analysis among the proposed state parameterization based dynamic output feedback controller \eqref{u optimal MF}, the state reconstruction based dynamic output feedback controller \eqref{u state reconstruction}, the Luenberger observer state parameterization based dynamic output feedback controller \eqref{u state parameterization}, and the static output feedback controller is given in this subsection.

According to \eqref{state reconstruction} and Theorem \ref{theorem control design}, note that both the optimal controllers \eqref{u state reconstruction} and \eqref{u optimal MF} are equivalent to the following full-state feedback controller
\begin{flalign}\label{u_full_state}
	u_x(k)=-{K^*} x(k)
\end{flalign}
where $K^*$ is obtained by \eqref{ARE K}. The controller \eqref{u state parameterization} equals to a traditional Luenberger state observer based output feedback controller
\begin{flalign}\label{u Luenberger observer}
    u_{\hat x}(k) = -K^* \hat x(k)
\end{flalign}
where $\hat x(k)$ is the estimated state by the Luenberger state observer \eqref{Luenberger observer}. This implies that the optimality analysis among controllers \eqref{u state reconstruction}, \eqref{u state parameterization}, and \eqref{u optimal MF} can be given by the analysis between the full-state feedback controller \eqref{u_full_state} and the Luenberger state observer based output feedback controller \eqref{u Luenberger observer}.

%Due to this equivalence relationship, the optimality of the controllers \eqref{controller MF} is guaranteed. 

\begin{proposition}\label{optimality proposition}
Given any weight matrices $Q_y = Q_y^\T \geq 0$ and $R = R^\T > 0$, the inequality
\begin{flalign*}
V_{u_{\hat x}}(x(k_0), \varepsilon_x(k_0) ) \geq V_{u_x} (x(k_0)), ~\forall x(k_0)\neq  \mathbf{0}, ~\varepsilon_x (k_0) \neq \mathbf{0}
\end{flalign*}
holds, where $V_{u_x} (x(k_0))$ and $V_{u_{\hat x}}(x(k_0), \varepsilon_x(k_0) )$ denote the value function \eqref{value_function} at an initial time instant $k_0$ under the state feedback controller \eqref{u_full_state} and the output feedback controller \eqref{u Luenberger observer}, respectively.  
\end{proposition}

\emph{Proof.} The closed-loop system with controller $u_x(k)=-K^* x(k)$ can be described as
\begin{flalign}\label{sys_closed_lemma2_full_state}
 x(k+1) = (A-B{K^*}) x(k).
\end{flalign}
The value function in \eqref{value_function} of the closed-loop system \eqref{sys_closed_lemma2_full_state} can be rewritten as 
\begin{flalign*}
	V_{u_x} (x(k_0)) = \sum_{k=k_0}^\infty x^\T(k) \left( Q_x + ({K^*})^\T R {K^*} \right) x(k) = x^\T(k_0) P_{u_x} x(k_0)
\end{flalign*}
where $P_{u_x}$ is the positive semi-definite solution to
\begin{flalign}\label{ARE K P}
    (A\!-\!BK^*)^\T P_{u_x}  (A\!-\!BK^*) \!-\!P_{u_x}  \!+\! C^\T Q_y C \!+\! ({K^*})^\T R {K^*} \!=\! \mathbf{0}.
\end{flalign} 

Define $\bar x = \col(x,\varepsilon_x)$. The closed-loop system with controller $u_{\hat x}(k)=-K^* \hat x(k)$ can be described as
\begin{flalign}\label{sys_closed_lemma2_OPFB}
	{\bar x} (k+1) \!=\! \left[ \begin{matrix}
		A-B{K^*} & B{K^*} \\
		0 & A-LC
	\end{matrix} \right] \bar x (k).
\end{flalign}
The utility function \eqref{cost_function} at the instant $k$ under control policy $u_{\hat x}(k)$ is given as
\begin{flalign} %\label{lemma2 eq6}
	&\mathcal{C}_{u_{\hat x}}(x(k), \varepsilon_x(k) ) \nonumber \\
	=& x^\T(k) C^\T Q_y C x(k) \!+\! \bar x^\T(k) \left[  \begin{matrix}
		({K^*})^\T R {K^*} & \!- ({K^*})^\T R {K^*} \\
		-({K^*})^\T R {K^*} & ({K^*})^\T R {K^*}
	\end{matrix} \right] \bar x(k) \nonumber \\
	=& \bar x^\T(k) \left[  \begin{matrix}
		C^\T Q_y C +({K^*})^\T R {K^*} & - ({K^*})^\T R {K^*} \\
		-({K^*})^\T R {K^*} & ({K^*})^\T R {K^*}
	\end{matrix} \right] \! \bar x(k). \nonumber
\end{flalign}
Then, the value function can be described by 
$$
V_{u_{\hat x}}(x(k_0), \varepsilon_x(k_0) ) = \bar x^\T(k_0) \left[  \begin{matrix}
		P_{11} & P_{12} \\
		P_{21} & P_{22}
	\end{matrix} \right] \bar x(k_0) :=\bar x^\T(k_0) P_{\hat u_x} \bar x(k_0).
$$
Thus, the Bellman equation can be described as
\begin{flalign}
    &\bar x^\T(k) P_{\hat u} \bar x(k)  \nonumber \\
    =& \bar x^\T (k\!+\!1) P_{\hat u}  \bar x(k\!+\!1) \nonumber \\
    &+ \bar x^\T(k) \left[  \begin{matrix}
		C^\T Q_y C +({K^*})^\T R {K^*} & - ({K^*})^\T R {K^*} \\
		-({K^*})^\T R {K^*} & ({K^*})^\T R {K^*}
	\end{matrix} \right] \! \bar x(k)   \label{bellman bar x}
\end{flalign}

Substituting \eqref{sys_closed_lemma2_OPFB} into \eqref{bellman bar x}, we have
\begin{subequations}
\begin{flalign}
\label{sub1} &(A\!-\!BK^*)^\T P_{11}  (A\!-\!BK^*) \!-\!P_{11} \!=\! - C^\T Q_y C \!-\! {({K^*}) }^\T{R}{ {K^*}} \\
\label{sub2} &{ (A \!-\! B{K^*}) }^\T {P_{11 }}B K^* \!+\! { (A \!-\! B{K^*}) }^\T P_{12} { (A \!-\! LC) } \!-\! {P_{12} } \!=\! {({K^*}) }^\T{R}{ {K^*}} \\
\label{sub3} &{ (A \!-\! LC) }^\T {P_{21 }}{ (A \!-\! B{K^*}) } \!+\! (B{K^*})^\T {P_{11} } { (A \!-\! B{K^*}) } \!-\! {P_{21} }\!=\! {({K^*}) }^\T{R}{ {K^*}} \\
\label{sub4} &
 (B{K^*})^\T {P_{11} } B{K^*} \!+\! { (A \!-\! LC) }^\T {P_{21 }} B K^* \!+\!(B{K^*})^\T {P_{12} } { (A \!-\! LC) }  \nonumber \\
 & +\! { (A \!-\! LC) }^\T P_{22} { (A \!-\! LC) } \!-\! P_{22} =\! -\!{({K^*}) }^\T{R}{ {K^*}}.
 \end{flalign}
\end{subequations}
It follows from \eqref{ARE K P} and \eqref{sub1} that 
\begin{equation}\label{Prop-eqp}
P_{11}=P_{u_x}
\end{equation}
%$P_{11}=P_{u_x}$. 
Note that $ P_{u_x}$ is a unique solution to the discrete algebraic Riccati equation \eqref{ARE}. From \eqref{ARE K} and \eqref{Prop-eqp}, we have ${ (A \!-\! B{K^*}) }^\T {P_{11 }}B K^* = (B{K^*})^\T {P_{11} } { (A \!-\! B{K^*}) } = \left( {K^*} \right)^\T{R}{ {K^*}}$. Thus, \eqref{sub2} and \eqref{sub3} can be rewritten as
\begin{flalign}
\label{sub2_re} & { (A \!-\! B{K^*}) }^\T P_{12} { (A \!-\! LC) } \!-\! {P_{12} } ~= \mathbf{0} \\
\label{sub3_re} &{ (A \!-\! LC) }^\T {P_{21 }}{ (A \!-\! B{K^*}) } -P_{21} = \mathbf{0}.
\end{flalign}
With the operators defined in Subsection \ref{notations}, \eqref{sub2_re} can be expressed as
\begin{flalign}\label{sub2_re_re}
    \left( (A \!-\! LC)^\T \otimes (A \!-\! B{K^*})^\T - I_n \otimes I_n \right) \mbox{vec}(P_{12}) = \mathbf{0}.
\end{flalign}
Since $\sigma(A \!-\! B{K^*}) \in \mathbb{U}^-$ and $\sigma(A \!-\! LC) \in \mathbb{U}^-$, the unique solution to \eqref{sub2_re_re} is $P_{12} = \mathbf{0}$. In the same way, we have $P_{21} = \mathbf{0}$. Then, \eqref{sub4} can be rewritten as
\begin{flalign}\label{sub4_re}  
&{ (A \!-\! LC) }^\T P_{22} { (A \!-\! LC) } \!-\! P_{22} \!=\! -{({K^*}) }^\T(R \!+\! B^\T P_{11} B){ {K^*}}\!.\!
\end{flalign}
Since $\sigma(A \!-\! LC) \in \mathbb{U}^-$, there always exists a unique positive semi-definite solution $P_{22}$ to \eqref{sub4_re}.
Thus, by \eqref{Prop-eqp}, the value function $V_{u_{\hat x}}(x(k_0), \varepsilon_x(k_0) )$ is given as
\begin{flalign*}
	V_{u_{\hat x}}(x(k_0), \varepsilon_x(k_0) ) = &x^\T (k_0)P_{11} x(k_0) \!+\! \varepsilon _x ^\T (k_0) P_{22} \varepsilon _x(k_0) \nonumber \\
                                          \geq & V_{u_x} (x(k_0)), ~~\forall x(k_0) \neq \mathbf{0}, \varepsilon_x(k_0) \neq \mathbf{0}.
\end{flalign*}
%It implies that the classical Luenberger observer-based output feedback controller \eqref{u Luenberger observer} is a sub-optimal control policy with respect to the pre-defined utility function \eqref{cost_function}. 
The proof is thus completed. \hfill \rule{1.5mm}{1.5mm}

Proposition \ref{optimality proposition} shows that the dynamic output feedback controllers \eqref{u state reconstruction} and \eqref{u optimal MF} can solve Problem \ref{problem 1}, while the controller \eqref{u state parameterization} is not a solution to Problem \ref{problem 1}.
The key step of the controller design of \eqref{u state reconstruction} and \eqref{u optimal MF} is to establish the equivalent relationship between the dynamic output feedback controllers and the full-state feedback controller.
The equivalent relationship guarantees the optimality of the dynamic output feedback controller to solve Problem \ref{problem 1}. 

Distinct from the dynamic output feedback controller, the static output feedback controller described by
\begin{flalign}
    \label{static OPFB controller}
    u_y(k) = -K_y y(k)
\end{flalign}
where $K_y$ is the feedback control gain, only needs to access the output data at the current instant. Compared to the proposed dynamic output feedback controller \eqref{u optimal MF}, the static output feedback controller may not have the ability to solve Problem \ref{problem 1}. 
We next give the optimality analysis between the dynamic output feedback controller and the static output feedback controller. 

To guarantee the feasibility of the static output feedback to the LQR problem, 
we assume that there exist real matrices $K_{y}$ and $H$ satisfying \citep{rosinova2003necessary}
\begin{flalign}\label{K_static_OPFB}
    K_{y } C - (R + B^\T P_{u_y} B)^{-1} B^\T P_{u_y} A = H 
\end{flalign}
where $P_{u_y}$ is the symmetric positive semi-definite solution to
\begin{flalign} \label{ARE static OPFB}
    \mathbf{0} =& A^\T P_{u_y} A \!-\! P_{u_y} \!-\! A^\T P_{u_y} B \left( R \!+\! B^\T  P_{u_y} B \right)^{\!-\!1} B^\T  P A \!+\! C^\T Q_y C \nonumber \\
    & \!+\! H^\T (R + B^\T P_{u_y} B) H.
\end{flalign}
The above assumption is a necessary and sufficient condition to ensure the stabilizability of the closed-loop system \eqref{sys_origin} under the static output feedback controller \eqref{static OPFB controller}. 
In what follows, we show that the static output feedback control may also be sub-optimal with respect to the utility function \eqref{cost_function}.

\begin{proposition}\label{optimality proposition 2} 
Given any weight matrices $Q_y = Q_y^\T \geq 0$ and $R = R^\T > 0$, the inequality
\begin{flalign}\label{app1_proposition_eq}
V_{u_y}(x(k_0) ) \geq V_{u_x} (x(k_0) ), ~~~~~\forall x(k_0)\neq  \mathbf{0}
\end{flalign}
holds, where $V_{u_x}(x(k_0))$ and $V_{u_y}(x(k_0))$ represent the value function \eqref{value_function} under the state feedback controller \eqref{u_full_state} and the static output feedback controller \eqref{static OPFB controller}, respectively. 
\end{proposition}

\emph{Proof.} Let $K'_y = (R + B^\T P_{u_y} B)^{-1} B^\T P_{u_y} A$ and $A_y = A-BK'_y$. 
It follows from \eqref{ARE static OPFB} that
\begin{flalign}\label{app1_eq1}
    \mathbf{0} \!=\! A_y^\T P_{u_y} A_y \!-\! P_{u_y} \!+\! {K'_y}^\T R  K'_y \!+\! C^\T Q_y C +\! H^\T (R \!+\! B^\T P_{u_y} B) H
\end{flalign}
Define $\mathcal{P}=A_y^\T P_{u_x} A_y - P_{u_x} + {K'_y}^\T R K'_y$, where $P_{u_x}$ is the solution to \eqref{ARE K P}. 
One can obtain
\begin{flalign}\label{app1_eq2}
    \mathcal{P} 
   \!=&(A-BK^*)^\T P_{u_x} (A-BK^*) -P_{u_x} + (K^*)^\T R K^* \nonumber \\
    & + \left({K^*} - K'_y \right)^\T (R + B^\T P_{u_x} B) \left({K^*} - K'_y \right) \nonumber \\
    =&-\!C^\T Q_y C \!+\! \left({K^*} \!-\! K'_y \right)^\T (R \!+\! B^\T P_{u_x} B) \left({K^*} \!-\! K'_y \right)
\end{flalign}
Subtracting \eqref{app1_eq1} from \eqref{app1_eq2}, we have
\begin{flalign}\label{app1_eq3}
    &A_y^\T (P_{u_y}-P_{u_x})A_y - (P_{u_y}-P_{u_x})  \\
    =& -\! H^\T (R \!+\! B^\T P_{u_y} B) H \!-\! \left({K^*} \!-\! K'_y \right)^\T (R \!+\! B^\T P_{u_x} B) \left({K^*} \!-\! K'_y \right). \nonumber 
\end{flalign}
Since $\sigma (A_y) \in \mathbb{U}^-$, $R > 0$, $P_{u_y} \geq 0$, and $P_{u_x}\geq 0$, the solution $(P_{u_y} - P_{u_x})$ to \eqref{app1_eq3} has the property that $ (P_{u_y} - P_{u_x}) \geq 0$.
Thus, we have 
\begin{flalign*}
V_{u_y}(x(k_0)) = x^\T (k_0) P_{u_y} x(k_0) \geq V_{u_x}(x(k_0)) = x^\T (k_0) P_{u_x} x(k_0)    
\end{flalign*}
This completes the proof.  \hfill \rule{1.5mm}{1.5mm}

%\begin{remark} \label{OPFB sub-optimal}
%It is easy to find that the existence of observer error is the essential reason leading to the sub-optimality of the classical Luenberger observer-based output feedback. For the static output feedback, if $C$ is of full column rank, for any $x(0)$, $y=Cx$ is a bijective function which means that the complete knowledge of the state $x(k)$ can be fed back to the closed-loop systems with a unique $y(k)$. Otherwise, the mapping from a series of $y(k)$ into $x(k)$ cannot be guaranteed to be bijective. Thus, the non-invertible matrix $C$ not only brings a theoretical challenge to solving \eqref{K_static_OPFB}--\eqref{ARE static OPFB} but also leads to the sub-optimal transient performance due to the non-zero pseudo-inverse error.
%\end{remark}

Finally, we give a generalized dynamic output feedback learning control solution to Problem \ref{problem 1} of which the convergence, stability, and optimality can be guaranteed without any prior knowledge of system matrices.

\begin{theorem} \label{theorem_optimality}
Under Assumption \ref{ass_1}, if the rank condition \eqref{rank condition} is satisfied, the dynamic output feedback controller \eqref{controller MF}, where the control gain is designed by the learning Algorithm 3 or 4 or 5, can solve Problem \ref{problem 1} with guaranteed convergence, stability, and optimality performance.
\end{theorem}

\emph{Proof.} Under Assumption \ref{ass_1}, %without any prior knowledge of system matrices, 
the internal model pair $(\mathcal{G}_1, \mathcal{G}_2)$ can be user-defined and $\mathcal{G}_1$ can be designed to have $n$ distinct eigenvalues, which implies that $A_L$ has $n$ distinct eigenvalues. Combining Lemma \ref{lemma rank full for M_error} and the definition of $M$ in \eqref{M define}, it is easy to get that the rank condition \eqref{rank condition M} holds when $A_L$ has $n$ distinct eigenvalues. If the rank conditions \eqref{rank condition} and \eqref{rank condition M} are satisfied, it follows from Theorems \ref{theorem convergence analysis VI} and \ref{theorem convergence analysis PI} that the estimated control gain $\mathcal{K}_j$ by Algorithm 3 or 4 converges to the optimal value $\mathcal{K}^*$. The stability performance of Algorithm 4 holds during the whole learning process due to the inherent property of the PI scheme as shown in \cite{Hewer1971}. Combining Lemma \ref{corollary modified C} and Theorems \ref{theorem convergence analysis VI} and \ref{theorem convergence analysis PI}, the convergence and stability performance of Algorithm 5 can be obtained.
Moreover, according to Propositions \ref{optimality proposition} and \ref{optimality proposition 2}, the control policy \eqref{controller MF} where the control gain is estimated by Algorithm 3 or 4 or 5 can solve Problem \ref{problem 1} with optimality. The proof is thus completed. \hfill \rule{1.5mm}{1.5mm}

\section{Simulation}\label{sec-sim}
In this section, two numerical simulations are given to verify the effectiveness of the proposed output feedback learning control method. 

\subsection{Unstable and uncontrollable system}\label{sec-sim1}
Consider the system described by
\begin{align*}
    x(k+1) =& \left[ \begin{matrix}
        1 & 0.5 \\
        0   &  0.6
    \end{matrix} \right] x(k) + \left[ \begin{matrix}
        1 \\
        0
    \end{matrix} \right] u(k) \\
    y(k) =& \left[ \begin{matrix}
        1 & 1
    \end{matrix} \right] x(k).  
\end{align*}
The initial state is $x(0)=\col\left(0,  10\right)$. The eigenvalues of the open-loop system are $1$ and $0.6$, and the system is uncontrollable.  

It is shown in \cite{Rizvi2023TAC} that the convergence of the learning algorithms in \cite{rizvi2019DToutput} and \cite{Rizvi2023TAC} cannot be guaranteed when one of the eigenvalues of the observer matrix $A_L$ is equal to $0.6$. 

To illustrate the effectiveness of our proposed learning algorithm, we chose the eigenvalues of $A_L$ as $0.6$ and $0.95$ and chose the initial state of the internal model as $\eta(0)=\col\left( 
        0, 0, 0,  0, 5, -5\right)$, which gives the pair $(\mathcal{G}_1, \mathcal{G}_2)$ defined in \eqref{G define} as $\mathcal{G}_1 = \mbox{block diag}[\mathcal{A}, \mathcal{A}, \mathcal{A}_{\varepsilon}], ~
\mathcal{G}_2 = \mbox{block diag}[b, b, \mathbf{0}]$
with
\begin{flalign*}
\mathcal{A}_{\varepsilon} \!=\! \mathcal{A} \!=\! \left[
                 \begin{array}{ccc}
            0  &  1 \\
           -0.57  &  1.55
                 \end{array}
               \right], ~~ b\!=\! \left[
                 \begin{array}{ccc}
            0  \\
            1
                 \end{array}
               \right]
\end{flalign*}\noindent
and the state parameterization matrices
\begin{flalign*}
    &M_u=\left[ \begin{matrix}
          -0.6  &  1 \\
           0  &   0
    \end{matrix} \right], M_y = \left[ \begin{matrix}
        -0.03  & 0.05 \\
         0    &  0
    \end{matrix} \right] \\
    &M_{\varepsilon_x \leftarrow \eta_\varepsilon} = \left[ \begin{matrix}
        -0.2885 &  -0.2885 \\
    0.9744 &  -1.0256
    \end{matrix} \right].
\end{flalign*}
Then, we set the weight matrices $Q_y = 1$ and $R = 1$, which gives the optimal control gain ${\mathcal{K}}^*$ of dynamic output feedback controller computed as
\begin{flalign*}
    {\mathcal{K}}^* \!=\![-0.3708,~
    0.6180,
   -0.0185,~
    0.0309,~
    0.5020,
   -0.8944]
\end{flalign*}
and $\| \mathcal{P}^*\|=6.1046$.

\begin{figure}[t]
\vspace{0pt}
  \centering
  \includegraphics[width=\hsize]{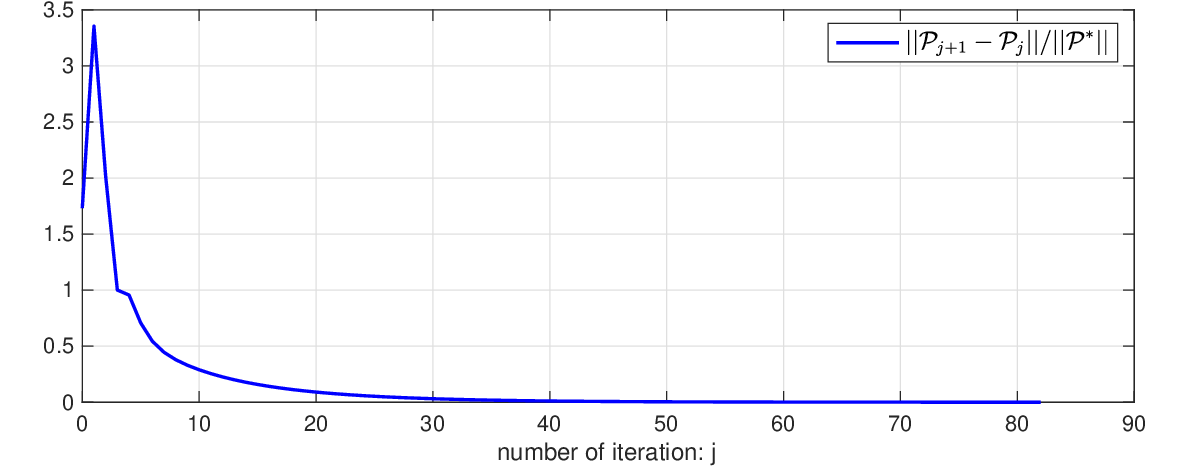}\vspace{-14pt}
  \caption{The convergence of Algorithm 3.}
  \label{fig-sim1-convergence}

\vspace{16pt}
  \centering
  \includegraphics[width=\hsize]{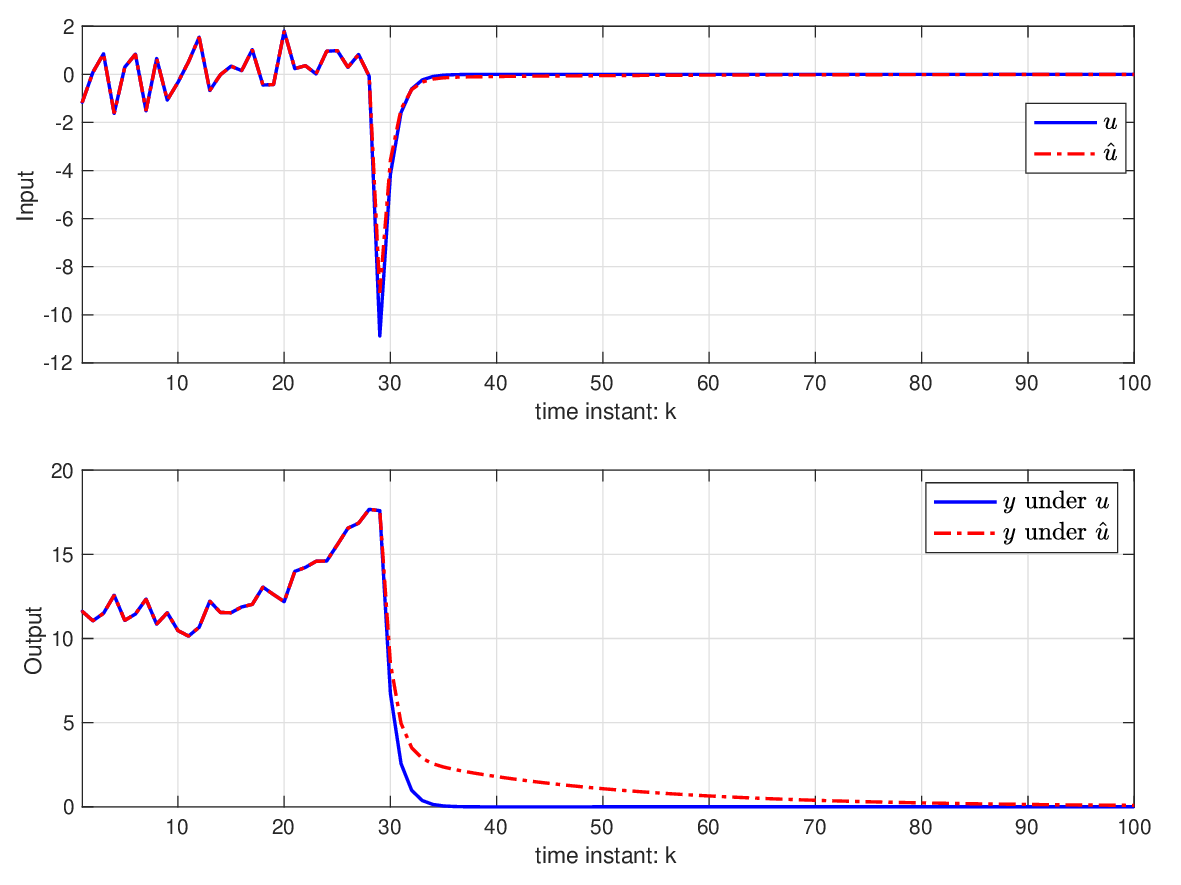}\vspace{-14pt}
  \caption{The trajectories of inputs and outputs under different controllers.}
  \label{fig-sim1-input output}

\vspace{16pt}
  \centering
  \includegraphics[width=\hsize]{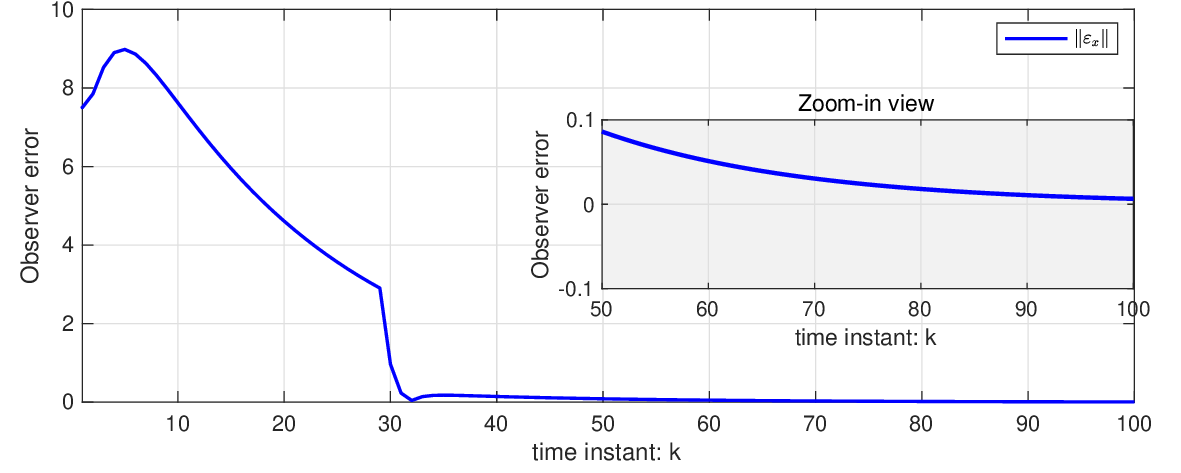}\vspace{-14pt}
  \caption{The evolution of observer error.}
  \label{fig-sim1-observer}
\end{figure}

The parameters in Algorithm 3 are chosen as $\mathcal{P}_0 = I_{n_z}$ with $n_z=6$, $\mathcal{K}_0 = \mathbf{0}$, and $\epsilon = 0.001$. The exploration noise is a combination of  $100$ sinusoids with random frequency $w_i$, amplitude $c_i$, and initial phase $d_i$, i.e., $\xi = \sum_{i=1}^{100} c_i \mbox{sin}(w_i k+ d_i) $. Two numerical simulations are carried out by employing Algorithm 3 with the proposed dynamic output feedback controller (denoted as $u$) and the traditional Luenberger observer-based output feedback controller (denoted as $\hat u$). The simulation results are shown in Figs. \ref{fig-sim1-convergence}--\ref{fig-sim1-observer}. 
Fig. \ref{fig-sim1-convergence} depicts the convergence performance of Algorithm 3, which shows that, after $83$ iterations, the estimated control gain is updated as
\begin{flalign*}
    {\tilde {\mathcal{K}}}^* =[-0.3708  ,~  0.6181,~   -0.0186  ,~  0.0309,~    0.5021 ,~  -0.8945].
\end{flalign*}
The evolution of inputs and outputs under the output feedback controllers $u$ and $\hat u$ is drawn in Fig. \ref{fig-sim1-input output}, which shows that the controllers are updated at $k_N=29$ when the rank condition is met. The evolution of the observer error is shown in Fig. \ref{fig-sim1-observer}. It follows from Figs. \ref{fig-sim1-input output} and \ref{fig-sim1-observer} that the transient performance of the closed-loop system under controller $u$ is better than that of the closed-loop system under controller $\hat u$.

\subsection{Application to the aircraft system}\label{sec-sim2}

Consider the aircraft system dynamics \citep{rizvi2019DToutput} linearized as
\begin{flalign*}
\left[ \begin{matrix}
                           \alpha(k\!+\!1)\\
                           q(k\!+\!1)\\
                           \delta_e (k\!+\!1)
\end{matrix} \right]  \!=& \left[ \begin{matrix}
                         0.906488 & 0.0816012 & -0.0005\\
   0.0741349 & 0.90121 & -0.0007083\\
   0         & 0  &     0.132655
\end{matrix} \right] \left[ \begin{matrix}
                            \alpha(k)\\
                               q(k)\\
                            \delta_e(k)
\end{matrix} \right] \\
&\!+\!
\left[ \begin{matrix}
                           -0.00150808\\
                           -0.0096\\
                           0.867345
\end{matrix} \right] \! u(k)  
\end{flalign*}
where $\alpha$, $q$, and ${\delta}_e$ denote the angle of attack, the pitch rate, and the elevator deflection angle, respectively. $u$ is the control input. $y = \alpha$ is the output. The initial state is $\col(    \alpha(0),   q(0),   \delta_e(0))= \col\left(
    0.2, 0.2,  0.2 \right)$.

Now, we choose the eigenvalues of the observer matrix $A_L$ as $-0.91$, $-0.92$, and $-0.93$, and select the initial state of internal model as $\eta(0)=\col\left(
        0,  0,  0,  0,  0,  0,  2,  1, 1 \right)$, which gives the pair $(\mathcal{G}_1, \mathcal{G}_2)$ defined in \eqref{G define} as $\mathcal{G}_1 = \mbox{block diag}[\mathcal{A}, \mathcal{A}, \mathcal{A}_{\varepsilon}], ~
\mathcal{G}_2 = \mbox{block diag}[b, b, \mathbf{0}]$
with
\begin{flalign*}
\mathcal{A}_{\varepsilon} &\!=\left[\! \begin{array}{ccc}
                -0.91 & 1.1 & -1.2\\
            0 & -0.92 & ~~~\!1.3\\
             0  & 0 &  -0.93
                 \end{array}
              \! \right] \\
 \mathcal{A} &\!= \left[\!
                 \begin{array}{ccc}
                0 & 1 & 0\\
            	0 & 0 & 1\\
             -0.7786  & -2.5391 &  -2.7600
                 \end{array}
              \! \right], ~~b \!=\! \left[\!
                 \begin{array}{ccc}
                0 \\
                0\\
                1
                 \end{array}
              \! \right]
\end{flalign*}
and the state parameterization matrices
\begin{flalign*}
    M_u &\!=\! \left[
                 \begin{array}{ccc}
    0.0003  &  0.0003  & -0.0015 \\
    0.0391  &  0.1145  & -0.0096 \\
    2.0276  &  7.8964  &  0.8673
                 \end{array}
               \right]  \\
   M_y &\!=\! \left[
                 \begin{array}{ccc}
0.8862 &	 1.4884	& 4.7004 \\
14.1018 &	-105.7656	& 99.6818\\
  2896.8322	& -6457.8672 &	3572.4259
                 \end{array}
               \right] \\
    M_{\varepsilon_x \leftarrow \eta_\varepsilon} &\!=\! \left[
                 \begin{array}{ccc}
    0.2505 &	-0.4134 &	0.1123 \\
14.1130 &	-14.2814 &	-13.7445 \\
858.3455 &	-505.9328 &	-1210.5582
                 \end{array}
               \right].
\end{flalign*}

\begin{figure}[!t]
\vspace{0pt}
  \centering
  \includegraphics[width=\hsize]{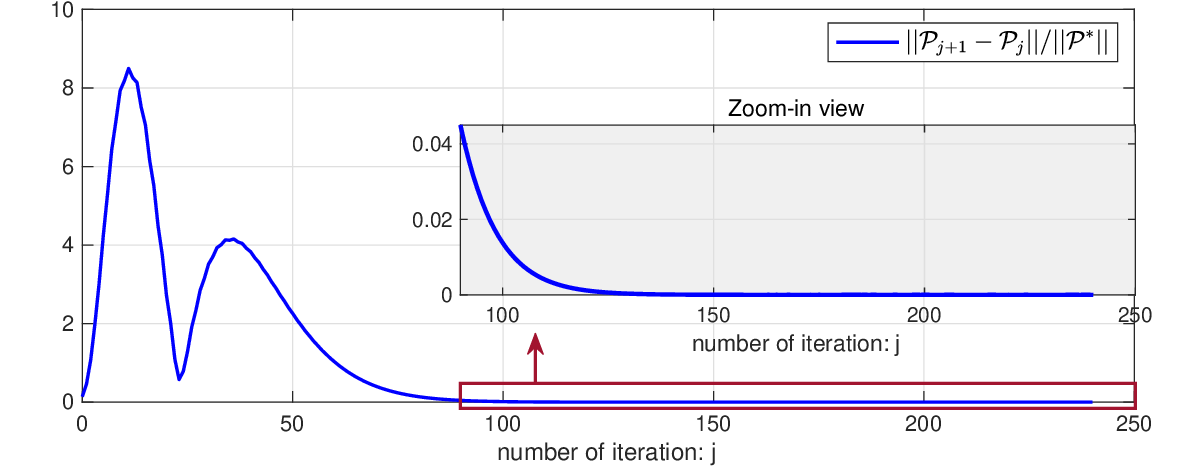}\vspace{-14pt}
  \caption{The convergence of VI-based Algorithm 3.}
  \label{fig-sim2-convergence}

\vspace{8pt}
  \centering
  \includegraphics[width=\hsize]{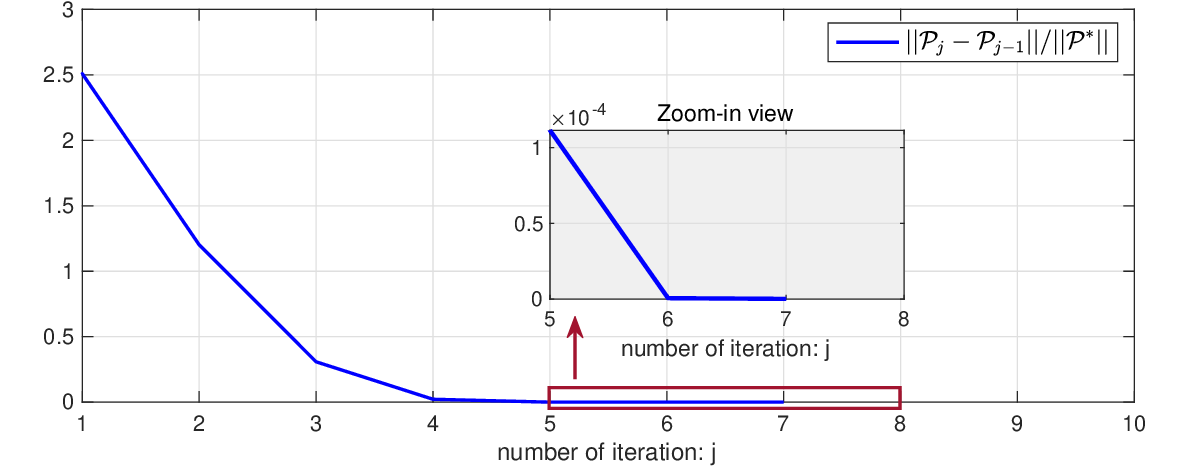}\vspace{-14pt}
  \caption{The convergence of PI-based Algorithm 4.}
  \label{fig-sim3-convergence}

\vspace{8pt}
  \centering
  \includegraphics[width=\hsize]{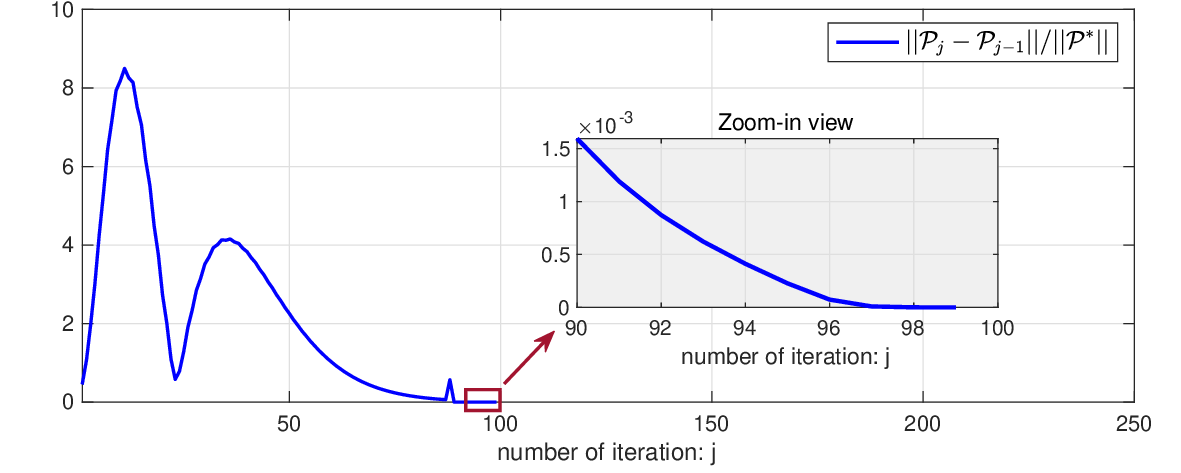}\vspace{-14pt}
  \caption{The convergence of SI-based Algorithm 5.}
  \label{fig-sim4-convergence}
\end{figure}

Then, we set the weight matrices $Q_y = 100$ and $R = 1$, which gives the optimal control gain ${\mathcal{K}}^*$ of dynamic output feedback controller computed as
\begin{flalign*}
    {\mathcal{K}}^* =[& -0.1029 ,~  -0.2900 ,~   0.0377,~  -30.0765,~  264.5137, \nonumber \\ & -286.4590 ,~ -37.1673 ,~  39.9946 ,~  32.9448]
\end{flalign*}
and $\| \mathcal{P}^*\|=4.2444 \times 10^6$.

\begin{figure}[!t]
\vspace{0pt}
  \centering
  \includegraphics[width=\hsize]{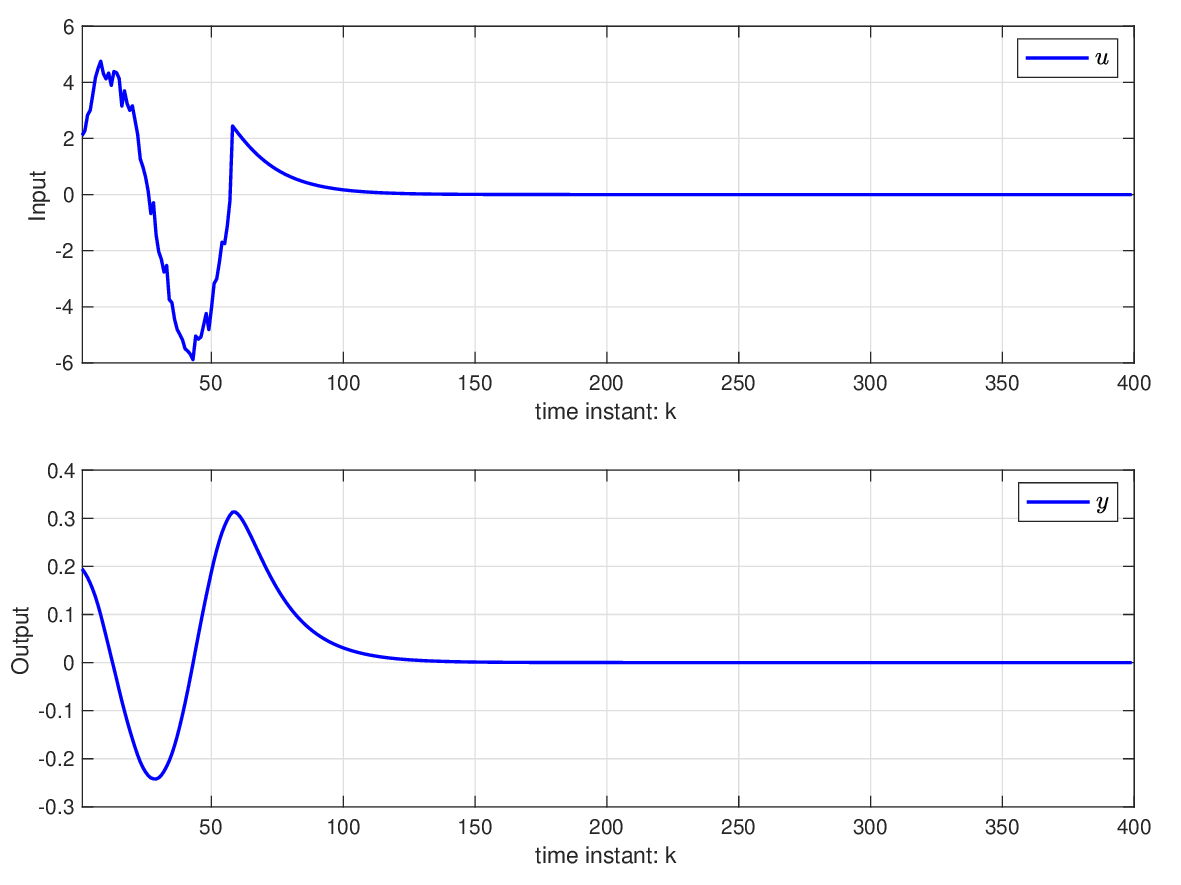}\vspace{-14pt}
  \caption{The evolution of input and output.}
  \label{fig-sim2-input-output}

\vspace{8pt}
  \centering
  \includegraphics[width=\hsize]{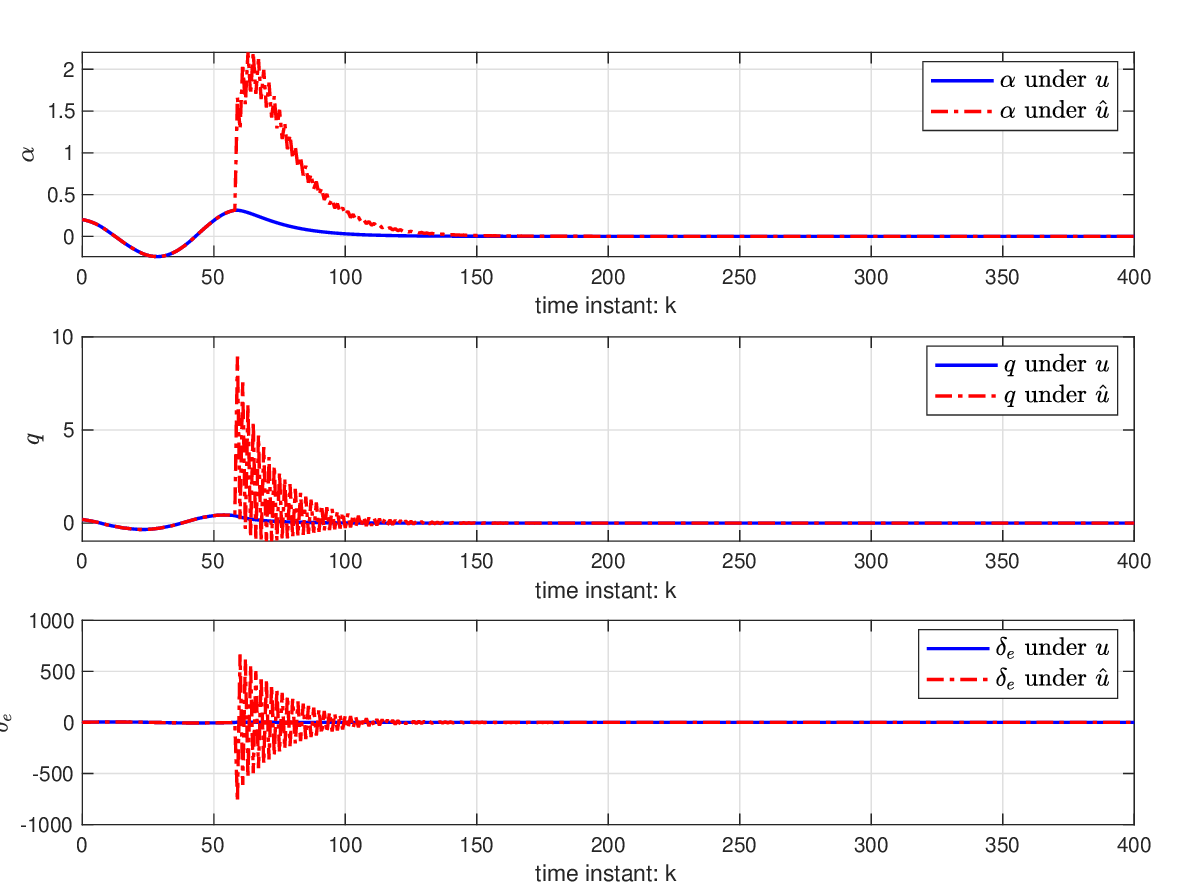}\vspace{-14pt}
  \caption{The evolution of states under different controllers.}
  \label{fig-sim2-states}

  \vspace{8pt}
  \centering
  \includegraphics[width=\hsize]{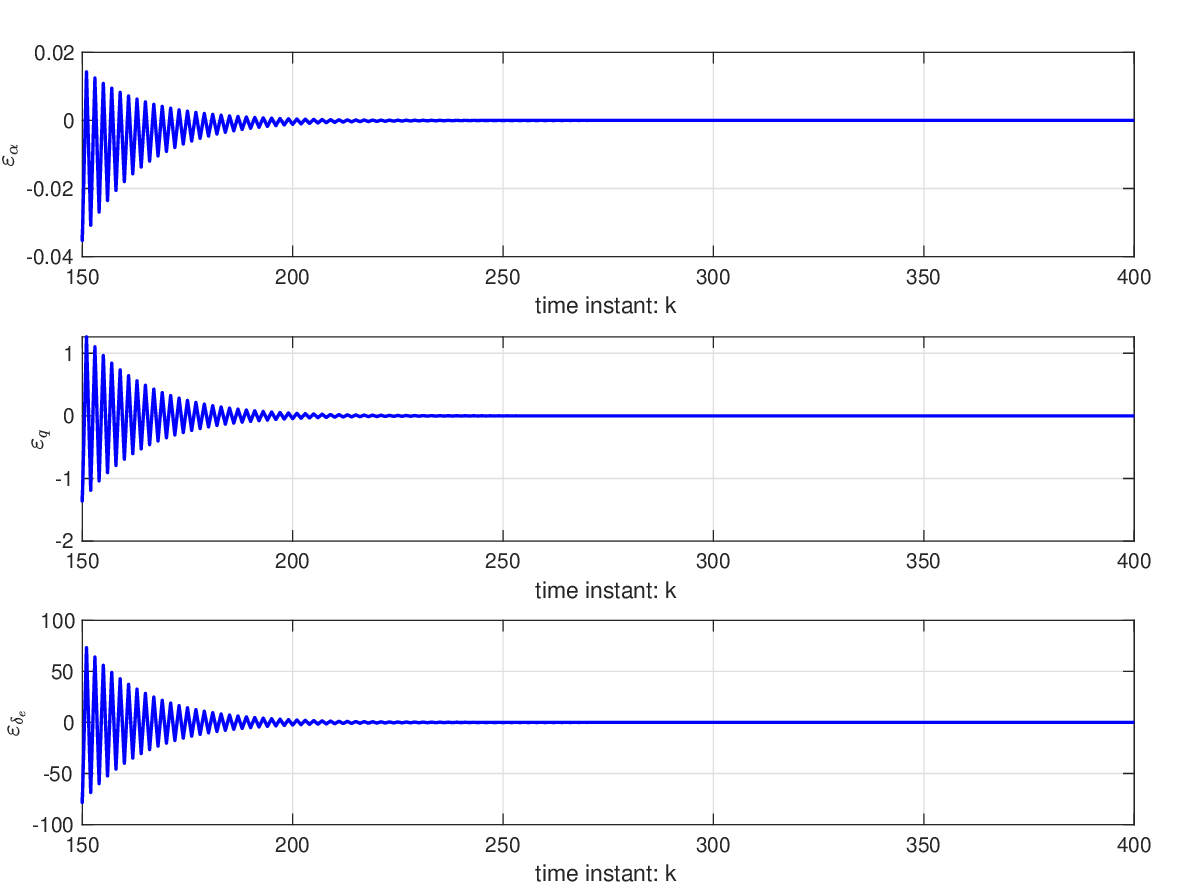}\vspace{-14pt}
  \caption{The evolution of the observer error.}
  \label{fig-sim2-observer-zoom}
\end{figure}

To illustrate the effectiveness of our proposed method, we made three numerical simulations by employing Algorithms 3, 4, and 5, respectively. 

The parameters are chosen as $\mathcal{P}_0 = 10^5 I_{n_z}$ with $n_z=9$, and $\epsilon = 1$. The exploration noise is a combination of $100$ sinusoids with random frequency $w_i$, amplitude $c_i$, and initial phase $d_i$, i.e., $\xi = \sum_{i=1}^{100} c_i \mbox{sin}(w_i k+ d_i) $. The admissible initial control gain is set as $\mathcal{K}_0 = \mathbf{0}$ in Algorithm 4, and we choose $\mathcal{K}_0^{\T} = \col(1.1069,~	3.3440, ~	-0.1596,~	469.8189,~	-3114.9114,~	2609.3162, ~ 399.8526,~	-382.6236,~	-418.7970)$ in Algorithms 3 and 5 which is not a stabilizing control gain. It should be noted that we set $Q_\varepsilon = 0.01 I_{n_z}$ in Algorithm 5. 
The simulation results are shown in Figs. \ref{fig-sim2-convergence}--\ref{fig-sim2-states}.
Fig. \ref{fig-sim2-convergence} depicts the convergence performance of VI-based Algorithm 3, which shows that, after $241$ iterations, the estimated control gain is updated as
\begin{flalign*}
    {\tilde {\mathcal{K}}}^* =[& -0.1033 ,~  -0.2910  ,~  0.0371 ,~ -30.0673,~  264.4322,~ \nonumber \\ & -286.3704,~  -37.1558 ,~  39.9822   ,~32.9344].
\end{flalign*}
It is shown in Fig. \ref{fig-sim3-convergence} that, for PI-based Algorithm 4, the estimated control gain is updated after only $7$ iterations, and its value is
\begin{flalign*}
    {\tilde {\mathcal{K}}}^* =[& -0.1027 ,~  -0.2896 ,~   0.0380 ,~ -30.0794 ,~ 264.5498,~ \nonumber \\ & -286.4994 ,~ -37.1724 ,~  40.0003  ,~ 32.9490].
\end{flalign*}
The convergence performance of SI-based Algorithm 5 is drawn in Fig. \ref{fig-sim4-convergence}, which shows that, after $99$ iterations, the estimated control gain is updated as
\begin{flalign*}
    {\tilde {\mathcal{K}}}^* =[& -0.1034 ,~  -0.2913  ,~  0.0369 ,~ -30.0650 ,~ 264.4091 ,~ \nonumber \\ & -286.3448,~  -37.1526 ,~  39.9786 ,~  32.9316].
\end{flalign*}
The main differences among Algorithms 3, 4, and 5 are twofold: 1) the convergence rate; and 2) the requirement of an initial admissible control policy. The above three numerical simulation results show that the SI-based Algorithm 5 is faster than the VI-based Algorithm 3, and it does not need the initial admissible control policy as the PI-based Algorithm 4 does.

In addition, similar to Subsection \ref{sec-sim1}, two numerical simulations are carried out by employing Algorithm 3 with $u$ and $\hat u$. 
The evolution of inputs and outputs of the closed-loop system is drawn in Fig. \ref{fig-sim2-input-output}, which shows that the controllers are updated at $k_N=58$ when the rank condition is met. 
The trajectory of states under the proposed dynamic output feedback controller $u$
and the traditional Luenberger observer-based output feedback
controller $\hat u$ are shown in Fig \ref{fig-sim2-states}. 
The evolution of observer error during time instant $k \in [150, 400]$ is shown in Fig. \ref{fig-sim2-observer-zoom}. 
From Figs. \ref{fig-sim2-input-output} -- \ref{fig-sim2-observer-zoom}, we can see that the proposed output feedback learning control approach is immune to the observer error, while the transient performance of the closed-loop systems under the Luenberger observer-based output feedback controller is influenced by the observer error.

\section{Conclusion}\label{sec-con}
This paper presents the optimal solution to the discrete-time LQR problem with unknown system matrices and unmeasurable states. 
The significance of this result is that a generalized dynamic output feedback learning control method is proposed. The model-free learning algorithms are established by employing the PI scheme, the VI scheme, and the SI scheme to estimate the optimal control policy to solve the LQR problem. 
In addition, we conducted in-depth discussions regarding the theoretical guarantee of the proposed ADP-based output feedback learning control methods, including convergence, stability, and optimality.

\scriptsize
\bibliographystyle{model5-names}        % Include this if you use bibtex
\bibliography{MyReference}

\begin{thebibliography}{34}
\expandafter\ifx\csname natexlab\endcsname\relax\def\natexlab#1{#1}\fi
\providecommand{\url}[1]{\texttt{#1}}
\providecommand{\href}[2]{#2}
\providecommand{\path}[1]{#1}
\providecommand{\DOIprefix}{doi:}
\providecommand{\ArXivprefix}{arXiv:}
\providecommand{\URLprefix}{URL: }
\providecommand{\Pubmedprefix}{pmid:}
\providecommand{\doi}[1]{\href{http://dx.doi.org/#1}{\path{#1}}}
\providecommand{\Pubmed}[1]{\href{pmid:#1}{\path{#1}}}
\providecommand{\bibinfo}[2]{#2}
\ifx\xfnm\relax \def\xfnm[#1]{\unskip,\space#1}\fi
%Type = Article
\bibitem[{Bian \& Jiang(2016)}]{Bian2016}
\bibinfo{author}{Bian, T.}, \& \bibinfo{author}{Jiang, Z.~P.} (\bibinfo{year}{2016}).
\newblock \bibinfo{title}{Value iteration and adaptive dynamic programming for data-driven adaptive optimal control design}.
\newblock {\it \bibinfo{journal}{Automatica}\/},  {\it \bibinfo{volume}{71}\/}, \bibinfo{pages}{348--360}.
%Type = Article
\bibitem[{Cao et~al.(2025)Cao, Su, Saddler, Cao, Wang, Lee, Siang, Luo, Pinchuk, Li et~al.}]{cao2025machine}
\bibinfo{author}{Cao, L.}, \bibinfo{author}{Su, J.}, \bibinfo{author}{Saddler, J.}, \bibinfo{author}{Cao, Y.}, \bibinfo{author}{Wang, Y.}, \bibinfo{author}{Lee, G.}, \bibinfo{author}{Siang, L.~C.}, \bibinfo{author}{Luo, Y.}, \bibinfo{author}{Pinchuk, R.}, \bibinfo{author}{Li, J.} et~al. (\bibinfo{year}{2025}).
\newblock \bibinfo{title}{Machine learning for real-time green carbon dioxide tracking in refinery processes}.
\newblock {\it \bibinfo{journal}{Renewable and Sustainable Energy Reviews}\/},  {\it \bibinfo{volume}{213}\/}, \bibinfo{pages}{115417}.
%Type = Article
\bibitem[{Chen et~al.(2023)Chen, Xie, Jiang, Xie \& Xie}]{Chen2023}
\bibinfo{author}{Chen, C.}, \bibinfo{author}{Xie, L.}, \bibinfo{author}{Jiang, Y.}, \bibinfo{author}{Xie, K.}, \& \bibinfo{author}{Xie, S.} (\bibinfo{year}{2023}).
\newblock \bibinfo{title}{Robust output regulation and reinforcement learning-based output tracking design for unknown linear discrete-time systems}.
\newblock {\it \bibinfo{journal}{IEEE Transactions on Automatic Control}\/},  {\it \bibinfo{volume}{68}\/}, \bibinfo{pages}{2391--2398}.
%Type = Article
\bibitem[{Chen et~al.(2022)Chen, Xie, Xie, Lewis \& Xie}]{Chen2022}
\bibinfo{author}{Chen, C.}, \bibinfo{author}{Xie, L.}, \bibinfo{author}{Xie, K.}, \bibinfo{author}{Lewis, L.~F.}, \& \bibinfo{author}{Xie, S.} (\bibinfo{year}{2022}).
\newblock \bibinfo{title}{Adaptive optimal output tracking of continuous-time systems via output-feedback-based reinforcement learning}.
\newblock {\it \bibinfo{journal}{Automatica}\/},  {\it \bibinfo{volume}{146}\/}, \bibinfo{pages}{110581}.
%Type = Book
\bibitem[{Chen(1999)}]{chen1984linear}
\bibinfo{author}{Chen, C.~T.} (\bibinfo{year}{1999}).
\newblock {\it \bibinfo{title}{Linear System Theory and Design, 3rd ed}\/}.
\newblock \bibinfo{publisher}{Oxford University Press}.
%Type = Article
\bibitem[{Dorato \& Levis(1971)}]{dorato1971optimal}
\bibinfo{author}{Dorato, P.}, \& \bibinfo{author}{Levis, A.} (\bibinfo{year}{1971}).
\newblock \bibinfo{title}{Optimal linear regulators: The discrete-time case}.
\newblock {\it \bibinfo{journal}{IEEE Transactions on Automatic Control}\/},  {\it \bibinfo{volume}{16}\/}, \bibinfo{pages}{613--620}.
%Type = Article
\bibitem[{Duan et~al.(2024)Duan, Cao, Zheng \& Zhao}]{Duan2023}
\bibinfo{author}{Duan, J.}, \bibinfo{author}{Cao, W.}, \bibinfo{author}{Zheng, Y.}, \& \bibinfo{author}{Zhao, L.} (\bibinfo{year}{2024}).
\newblock \bibinfo{title}{On the optimization landscape of dynamic output feedback linear quadratic control}.
\newblock {\it \bibinfo{journal}{IEEE Transactions on Automatic Control}\/},  {\it \bibinfo{volume}{69}\/}, \bibinfo{pages}{920--935}.
%Type = Article
\bibitem[{Gao et~al.(2016)Gao, Jiang, Jiang \& Chai}]{Gao2016-3aut}
\bibinfo{author}{Gao, W.}, \bibinfo{author}{Jiang, Y.}, \bibinfo{author}{Jiang, Z.~P.}, \& \bibinfo{author}{Chai, T.} (\bibinfo{year}{2016}).
\newblock \bibinfo{title}{Output-feedback adaptive optimal control of interconnected systems based on robust adaptive dynamic programming}.
\newblock {\it \bibinfo{journal}{Automatica}\/},  {\it \bibinfo{volume}{72}\/}, \bibinfo{pages}{37--45}.
%Type = Article
\bibitem[{Gao \& Jiang(2016)}]{Gao2016_1tac}
\bibinfo{author}{Gao, W.}, \& \bibinfo{author}{Jiang, Z.~P.} (\bibinfo{year}{2016}).
\newblock \bibinfo{title}{Adaptive dynamic programming and adaptive optimal output regulation of linear systems}.
\newblock {\it \bibinfo{journal}{IEEE Transactions on Automatic Control}\/},  {\it \bibinfo{volume}{61}\/}, \bibinfo{pages}{4164--4169}.
%Type = Article
\bibitem[{Gao \& Jiang(2019)}]{Gao2019}
\bibinfo{author}{Gao, W.}, \& \bibinfo{author}{Jiang, Z.~P.} (\bibinfo{year}{2019}).
\newblock \bibinfo{title}{Adaptive optimal output regulation of time-delay systems via measurement feedback}.
\newblock {\it \bibinfo{journal}{IEEE Transactions on Neural Networks and Learning Systems}\/},  {\it \bibinfo{volume}{30}\/}, \bibinfo{pages}{938--945}.
%Type = Article
\bibitem[{Hewer(1971)}]{Hewer1971}
\bibinfo{author}{Hewer, G.} (\bibinfo{year}{1971}).
\newblock \bibinfo{title}{An iterative technique for the computation of the steady state gains for the discrete optimal regulator}.
\newblock {\it \bibinfo{journal}{IEEE Transactions on Automatic Control}\/},  {\it \bibinfo{volume}{16}\/}, \bibinfo{pages}{382--384}.
%Type = Article
\bibitem[{Jiang \& Jiang(2012)}]{Jiang2012}
\bibinfo{author}{Jiang, Y.}, \& \bibinfo{author}{Jiang, Z.~P.} (\bibinfo{year}{2012}).
\newblock \bibinfo{title}{Computational adaptive optimal control for continuous-time linear systems with completely unknown dynamics}.
\newblock {\it \bibinfo{journal}{Automatica}\/},  {\it \bibinfo{volume}{48}\/}, \bibinfo{pages}{2699--2704}.
%Type = Book
\bibitem[{Jiang \& Jiang(2017)}]{Jiang2017_book}
\bibinfo{author}{Jiang, Y.}, \& \bibinfo{author}{Jiang, Z.~P.} (\bibinfo{year}{2017}).
\newblock {\it \bibinfo{title}{Robust Adaptive Dynamic Programming}\/}.
\newblock \bibinfo{address}{Hoboken , NJ, USA}: \bibinfo{publisher}{Wiley}.
%Type = Article
\bibitem[{Kalman et~al.(1960)}]{kalman1960contributions}
\bibinfo{author}{Kalman, R.~E.} et~al. (\bibinfo{year}{1960}).
\newblock \bibinfo{title}{Contributions to the theory of optimal control}.
\newblock {\it \bibinfo{journal}{Bol. soc. mat. mexicana}\/},  {\it \bibinfo{volume}{5}\/}, \bibinfo{pages}{102--119}.
%Type = Book
\bibitem[{Lancaster \& Rodman(1995)}]{Lancaster1995}
\bibinfo{author}{Lancaster, P.}, \& \bibinfo{author}{Rodman, L.} (\bibinfo{year}{1995}).
\newblock {\it \bibinfo{title}{Algebraic Riccati Equations}\/}.
\newblock \bibinfo{address}{New York, NY, USA}: \bibinfo{publisher}{Clarendon press}.
%Type = Article
\bibitem[{Lewis \& Vamvoudakis(2011)}]{Lewis2011}
\bibinfo{author}{Lewis, F.~L.}, \& \bibinfo{author}{Vamvoudakis, K.~G.} (\bibinfo{year}{2011}).
\newblock \bibinfo{title}{Reinforcement learning for partially observable dynamic processes: Adaptive dynamic programming using measured output data}.
\newblock {\it \bibinfo{journal}{IEEE Transactions on Systems, Man, and Cybernetics, Part B (Cybernetics)}\/},  {\it \bibinfo{volume}{41}\/}, \bibinfo{pages}{14--25}.
%Type = Book
\bibitem[{Lewis et~al.(2012{\natexlab{a}})Lewis, Vrabie \& Syrmos}]{Lewis2012}
\bibinfo{author}{Lewis, F.~L.}, \bibinfo{author}{Vrabie, D.~L.}, \& \bibinfo{author}{Syrmos, V.~L.} (\bibinfo{year}{2012}{\natexlab{a}}).
\newblock {\it \bibinfo{title}{Optimal Control}\/}.
\newblock \bibinfo{publisher}{John Wiley \& Sons}.
%Type = Article
\bibitem[{Lewis et~al.(2012{\natexlab{b}})Lewis, Vrabie \& Vamvoudakis}]{Lewis2012introduction}
\bibinfo{author}{Lewis, F.~L.}, \bibinfo{author}{Vrabie, D.~L.}, \& \bibinfo{author}{Vamvoudakis, K.~G.} (\bibinfo{year}{2012}{\natexlab{b}}).
\newblock \bibinfo{title}{Reinforcement learning and feedback control: Using natural decision methods to design optimal adaptive controllers}.
\newblock {\it \bibinfo{journal}{IEEE Control Systems Magazine}\/},  {\it \bibinfo{volume}{32}\/}, \bibinfo{pages}{76--105}.
%Type = Article
\bibitem[{{Liu} et~al.(2021){Liu}, {Xue}, {Zhao}, {Luo} \& {Wei}}]{Liu2021}
\bibinfo{author}{{Liu}, D.}, \bibinfo{author}{{Xue}, S.}, \bibinfo{author}{{Zhao}, B.}, \bibinfo{author}{{Luo}, B.}, \& \bibinfo{author}{{Wei}, Q.} (\bibinfo{year}{2021}).
\newblock \bibinfo{title}{Adaptive dynamic programming for control: A survey and recent advances}.
\newblock {\it \bibinfo{journal}{IEEE Transactions on System Man Cybernetics: System}\/},  {\it \bibinfo{volume}{51}\/}, \bibinfo{pages}{142--160}.
%Type = Article
\bibitem[{Modares \& Lewis(2014)}]{Modares2014}
\bibinfo{author}{Modares, H.}, \& \bibinfo{author}{Lewis, F.~L.} (\bibinfo{year}{2014}).
\newblock \bibinfo{title}{Linear quadratic tracking control of partially-unknown continuous-time systems using reinforcement learning}.
\newblock {\it \bibinfo{journal}{IEEE Transactions on Automatic Control}\/},  {\it \bibinfo{volume}{59}\/}, \bibinfo{pages}{3051--3056}.
%Type = Article
\bibitem[{Modares et~al.(2016)Modares, Lewis \& Jiang}]{Modares2016}
\bibinfo{author}{Modares, H.}, \bibinfo{author}{Lewis, F.~L.}, \& \bibinfo{author}{Jiang, Z.} (\bibinfo{year}{2016}).
\newblock \bibinfo{title}{Optimal output-feedback control of unknown continuous-time linear systems using off-policy reinforcement learning}.
\newblock {\it \bibinfo{journal}{IEEE Transactions on Cybernetics}\/},  {\it \bibinfo{volume}{46}\/}, \bibinfo{pages}{2401--2410}.
%Type = Article
\bibitem[{Odekunle et~al.(2020)Odekunle, Gao, Davari \& Jiang}]{Gao2020-NZS}
\bibinfo{author}{Odekunle, A.}, \bibinfo{author}{Gao, W.}, \bibinfo{author}{Davari, M.}, \& \bibinfo{author}{Jiang, Z.~P.} (\bibinfo{year}{2020}).
\newblock \bibinfo{title}{Reinforcement learning and non-zero-sum game output regulation for multi-player linear uncertain systems}.
\newblock {\it \bibinfo{journal}{Automatica}\/},  {\it \bibinfo{volume}{112}\/}, \bibinfo{pages}{108672}.
%Type = Article
\bibitem[{Pang et~al.(2022)Pang, Bian \& Jiang}]{Bian-2022-robust}
\bibinfo{author}{Pang, B.}, \bibinfo{author}{Bian, T.}, \& \bibinfo{author}{Jiang, Z.~P.} (\bibinfo{year}{2022}).
\newblock \bibinfo{title}{Robust policy iteration for continuous-time linear quadratic regulation}.
\newblock {\it \bibinfo{journal}{IEEE Transactions on Automatic Control}\/},  {\it \bibinfo{volume}{67}\/}, \bibinfo{pages}{504--511}.
%Type = Article
\bibitem[{Postoyan et~al.(2016)Postoyan, Busoniu, Nesic \& Daafouz}]{postoyan2016stability}
\bibinfo{author}{Postoyan, R.}, \bibinfo{author}{Busoniu, L.}, \bibinfo{author}{Nesic, D.}, \& \bibinfo{author}{Daafouz, J.} (\bibinfo{year}{2016}).
\newblock \bibinfo{title}{Stability analysis of discrete-time infinite-horizon optimal control with discounted cost}.
\newblock {\it \bibinfo{journal}{IEEE Transactions on Automatic Control}\/},  {\it \bibinfo{volume}{62}\/}, \bibinfo{pages}{2736--2749}.
%Type = Book
\bibitem[{Powell(2004)}]{Powell2007}
\bibinfo{author}{Powell, W.~B.} (\bibinfo{year}{2004}).
\newblock {\it \bibinfo{title}{Approximate Dynamic Programming: Solving the Curse of Dimensionality}\/}.
\newblock \bibinfo{address}{New York, NY, USA}: \bibinfo{publisher}{Wiley}.
%Type = Article
\bibitem[{Rizvi \& Lin(2019)}]{rizvi2019DToutput}
\bibinfo{author}{Rizvi, S. A.~A.}, \& \bibinfo{author}{Lin, Z.} (\bibinfo{year}{2019}).
\newblock \bibinfo{title}{Output feedback q-learning control for the discrete-time linear quadratic regulator problem}.
\newblock {\it \bibinfo{journal}{IEEE Transactions on Neural Networks and Learning Systems}\/},  {\it \bibinfo{volume}{30}\/}, \bibinfo{pages}{1523--1536}.
%Type = Article
\bibitem[{Rizvi \& Lin(2020)}]{Rizvi2020_aut}
\bibinfo{author}{Rizvi, S. A.~A.}, \& \bibinfo{author}{Lin, Z.} (\bibinfo{year}{2020}).
\newblock \bibinfo{title}{Output feedback adaptive dynamic programming for linear differential zero-sum games}.
\newblock {\it \bibinfo{journal}{Automatica}\/},  {\it \bibinfo{volume}{122}\/}, \bibinfo{pages}{109272}.
%Type = Article
\bibitem[{Rizvi \& Lin(2023)}]{Rizvi2023TAC}
\bibinfo{author}{Rizvi, S. A.~A.}, \& \bibinfo{author}{Lin, Z.} (\bibinfo{year}{2023}).
\newblock \bibinfo{title}{A note on state parameterizations in output feedback reinforcement learning control of linear systems}.
\newblock {\it \bibinfo{journal}{IEEE Transactions on Automatic Control}\/},  {\it \bibinfo{volume}{68}\/}, \bibinfo{pages}{6200--6207}.
%Type = Article
\bibitem[{Rosinov{\'a} et~al.(2003)Rosinov{\'a}, Vesel{\`y} \& Ku{$\check{\mbox{c}}$}era}]{rosinova2003necessary}
\bibinfo{author}{Rosinov{\'a}, D.}, \bibinfo{author}{Vesel{\`y}, V.}, \& \bibinfo{author}{Ku{$\check{\mbox{c}}$}era, V.} (\bibinfo{year}{2003}).
\newblock \bibinfo{title}{A necessary and sufficient condition for static output feedback stabilizability of linear discrete-time systems}.
\newblock {\it \bibinfo{journal}{Kybernetika}\/},  {\it \bibinfo{volume}{39}\/}, \bibinfo{pages}{447--459}.
%Type = Article
\bibitem[{Su et~al.(2020)Su, Chow, Zheng, Huang, Liang \& Zhong}]{su2020neuro}
\bibinfo{author}{Su, Z.}, \bibinfo{author}{Chow, A.~H.}, \bibinfo{author}{Zheng, N.}, \bibinfo{author}{Huang, Y.}, \bibinfo{author}{Liang, E.}, \& \bibinfo{author}{Zhong, R.} (\bibinfo{year}{2020}).
\newblock \bibinfo{title}{Neuro-dynamic programming for optimal control of macroscopic fundamental diagram systems}.
\newblock {\it \bibinfo{journal}{Transportation Research Part C: Emerging Technologies}\/},  {\it \bibinfo{volume}{116}\/}, \bibinfo{pages}{102628}.
%Type = Book
\bibitem[{Sutton \& Barto(2018)}]{Sutton2018}
\bibinfo{author}{Sutton, R.~S.}, \& \bibinfo{author}{Barto, A.~G.} (\bibinfo{year}{2018}).
\newblock {\it \bibinfo{title}{Reinforcement Learning\mbox{:} An Introduction}\/}.
\newblock \bibinfo{address}{Cambridge, MA, USA}: \bibinfo{publisher}{MIT Press}.
%Type = Article
\bibitem[{Vrabie et~al.(2009)Vrabie, Pastravanu, Abu-Khalaf \& Lewis}]{vrabie2009adaptive}
\bibinfo{author}{Vrabie, D.~L.}, \bibinfo{author}{Pastravanu, O.}, \bibinfo{author}{Abu-Khalaf, M.}, \& \bibinfo{author}{Lewis, F.~L.} (\bibinfo{year}{2009}).
\newblock \bibinfo{title}{Adaptive optimal control for continuous-time linear systems based on policy iteration}.
\newblock {\it \bibinfo{journal}{Automatica}\/},  {\it \bibinfo{volume}{45}\/}, \bibinfo{pages}{477--484}.
%Type = Article
\bibitem[{Wu et~al.(2025)Wu, Xiao \& Braatz}]{wu2025eiqp}
\bibinfo{author}{Wu, L.}, \bibinfo{author}{Xiao, W.}, \& \bibinfo{author}{Braatz, R.~D.} (\bibinfo{year}{2025}).
\newblock \bibinfo{title}{Eiqp: Execution-time-certified and infeasibility-detecting qp solver}.
\newblock {\it \bibinfo{journal}{arXiv preprint arXiv:2502.07738}\/}, .
%Type = Article
\bibitem[{Yang et~al.(2018)Yang, He \& Zhong}]{Yang2018_1}
\bibinfo{author}{Yang, X.}, \bibinfo{author}{He, H.}, \& \bibinfo{author}{Zhong, X.} (\bibinfo{year}{2018}).
\newblock \bibinfo{title}{Adaptive dynamic programming for robust regulation and its application to power systems}.
\newblock {\it \bibinfo{journal}{IEEE Transactions on Industrial Electronics}\/},  {\it \bibinfo{volume}{65}\/}, \bibinfo{pages}{5722--5732}.

\end{thebibliography}

\end{document}